\documentclass[review]{elsarticle}

\usepackage{graphicx}
\usepackage{amsmath}
\usepackage{CJK}
\usepackage{lineno,hyperref}
\modulolinenumbers[100]

\journal{Elsevier}

\begin{document}

\begin{frontmatter}

\title{Consistent implementation of characteristic flux-split based finite difference method for compressible multi-material flows}


\author{Zhiwei He}
\author{Yousheng Zhang}
\author{Li Li}
\author{Baolin Tian\corref{hoho}}
\cortext[hoho]{Corresponding author}
\ead{tian{\_}baolin@iapcm.ac.cn}

\address{Laboratory of Computational Physics, Institute of Applied Physics and Computational Mathematics, Beijing 100094, China}

\begin{abstract}
In order to prevent velocity, pressure, and temperature spikes at material discontinuities occurring when the interface-capturing schemes inconsistently simulate compressible multi-material flows(when the specific heats ratio is variable), various non-conservative or quasi-conservative numerical models have been proposed. However, designing a consistent numerical algorithm, especially using the high-order characteristic flux-split based finite-difference method (CFS-FDM), is still an open question. In this study, a systematical analysis of previous algorithms of the consistent implementing the high-order CFS-FDM for such flows is performed, and the reasons of special treatments in these algorithms are revealed. Based on this analysis, a new general numerical methodology that successfully avoids any special treatments as those required in previously reported algorithms, is derived. In this new algorithm, we rewrite the non-conservative term as a conservative term with a source term containing velocity divergence. By consistently treating the advection velocity in the conservative term and velocity divergence in the source term by imposing a new additional criterion, specifically, that a multi-fluid algorithm should have the ability of maintaining a pure single-fluid, we finally derive a new general algorithm that does not need any special treatment, and is very convenient to implement. The results of some benchmark tests show that the final algorithm not only maintains the velocity, pressure, and temperature equlibria, but is also suitable for problems regarding the interaction of interfaces and strong shock and rarefaction waves.
\end{abstract}


\end{frontmatter}

\linenumbers

\section{Introduction}

Compressible multi-material flows arise in various natural and industrial fields, such as in inertial confinement fusion, high-speed combustion, and supernova explosion. The numerical computation of such flows has continuously been an active research field.

Generally, there are two types of methods for studying such flows: sharp interface methods (SIM) that consider an interface as a sharp discontinuity, and diffuse interface methods (DIM) that consider an interface as a diffuse zone, such as contact discontinuities in gas dynamics \cite{Saurel2009}. The arbitrary Lagrangian--Eulerian methods, free-Lagrange methods, volume of fluid (VOF) approach, level set approach, and front tracking approach are typical examples of SIM. However, computing large interface deformations continues to be challenging for these methods.

In contrast, the diffuse interfaces in DIM are computed as artificial mixtures generated by a numerical diffusion. The same algorithm can be implemented globally for both pure fluids and mixture zones. This is the most attractive advantage for computing flows with large interface deformations, such as the flows associated with interfacial instability and/or turbulent mixing \cite{Brouillette2002RMI, Dimotakis2005}. For general multiphase flows, the recent developments in DIM for compressible multiphase flows originate from a two-fluid model \cite{Baer1986, Saurel1999JCP} in which the balance equations for mass, momentum, and energy of each phase, as well as the equation for the volume fraction evolution, are solved. Baer--Nunziato¡¯s model \cite{Baer1986} is the best known two-fluid model. In addition, comparatively simpler and more compact models have also been proposed and successfully applied \cite{Kreeft2010}. An elegant hierarchy of reduced models exists, with the numbers of equations ranging from three to six only \cite{Saurel2009, Kapila2001,Kreeft2010, Murrone2005, Allaire2002, hezw2017}. For multi-material gas flows such as in a thermodynamic state in which the specific heats ratio is different between two fluids, the multispecies model is still widely used, particularly for studying the Richtmyer--Meshkov instability and/or turbulent mixing \cite{Tritschler2014, Cook2009}.

This study is mainly concerned with multi-material gas flows (when the specific heats ratio is variable), and therefore, the multispecies model is used. However, a naive implementation of standard shock-capturing techniques for this model has long been known to give rise to spurious pressure oscillations at interfaces. Circumventing the generation of nonphysical oscillations at the interfaces of this model is an attractive area for numerous researchers. Karni \cite{Karni1994} introduced a non-conservative model using primitive variables to avoid the pressure oscillations that was later modified to capture strong shock waves using the pressure evolution equation. Abgrall \cite{Abgrall1996} pioneered the addressing of this nonphysical pressure oscillation problem, and solved multi-material gas flows by introducing an advection form for a given function of the specific heats ratio under the ideal gas law (commonly called the quasi-conservative form). Shyue \cite{Shyue1998} later extended this idea to solving the transport equation for the mass fraction in advection form as well. In order to further prevent the temperature oscillations, Johnsen and Ham \cite{Johnsen2012} proposed an overestimation formulation, while Beig and Johnsen \cite{Johnsen2015JCP} systematically analyzed the temperature oscillations phenomenon, and demonstrated that the material properties entering the equation of state must be computed according to suitable transport equations in conservative or non-conservative forms; the pressure and temperature must be calculated based on the appropriate properties.

All the formulations discussed above are non-conservative or quasi-conservative. The presence of non-conservative products poses a huge computational challenge. So far, Abgrall's approach has been realized within the framework of the finite-volume method (FVM) with first- and second-order variable reconstructions using various Riemann solvers \cite{Saurel&Abgrall1999} and with high-order WENO reconstruction through the HLLC solver \cite{Johnsen2006}.

However, as stated in \cite{Hezw2015}, the high-order finite-difference method (FDM) is more difficult to be implemented for such non-conservative numerical models. Till date, there are only a few results for compressible multi-material gas flows computed with the FDM. Marquina and Mulet \cite{Marquina2003} directly solved the conservation form of the governing equations with the FDM. Terashima et al. \cite{Terashima2013} directly implemented central finite difference schemes to simulate multi-material flows by introducing consistent local artificial diffusion terms to suppress the numerical oscillations of the pressure and velocity. Nonomura et al. \cite{Nonomura2012} selected the weighted-compact nonlinear scheme (WCNS) variable interpolation finite-difference formulation to take over the numerical technique developed with the FVM.  In \cite{Hezw2015}, we performed a systematic analysis of the computational difficulties of the component-wise flux-split based FDM, particularly for the frequently used nonlinear finite-difference WENO schemes \cite{Shu1996}, when solving problems of compressible multi-material gas flows. We mainly used the same weight technique in the WENO scheme for almost all the flux components. When the local characteristic decomposition was used, we subsequently found \cite{hezw2016} that the constraint of the above-mentioned same weight technique in a WENO scheme could be released. The velocity and pressure equilibria can be maintained if the nonlinear WENO schemes in the genuinely nonlinear characteristic fields can be ensured to be the same and the decomposition equation representing the material interfaces is appropriately discretized. Recently, Nonomura and Fujii \cite{Nonomura2017} proposed a new characteristic finite-difference WENO scheme for compressible multi-material gas flows that is expressed in a split form including the consistent and dissipation parts of the numerical flux (linear central scheme with an artificial diffusion term). The dissipation part of the numerical flux is modified in the conservative form to maintain the conservation of the conservative variables. The scheme implemented in their study can preserve the velocity, pressure, and temperature equilibria. However, this distinct treatment of the WENO scheme is still unsatisfactory.

It appears that there is still no general solution for the consistent implementation of the characteristic flux-split based FDM for compressible multi-material gas flows. In this paper, we present a detailed analysis of the problem, and propose a general solution without any special treatment. This work mainly contains
\begin{enumerate}[(1)]
  \item A systematical analysis of the application of the previous two methods \cite{hezw2016, Nonomura2017} for consistently implementing a high-order characteristic flux-split based FDM for multi-material gas flow is performed. The reasons for the use of a common discretization of genuinely nonlinear fields or the use of the split form of the nonlinear WENO scheme in these two methods are discussed in section 3.
  \item We convert the non-conservative term into a conservative form with a source term containing the velocity divergence, and propose a general framework that is found to have the ability to maintain the velocity, pressure, and temperature equilibria. Furthermore, the consistent discretization form of the velocity divergence in the source term can also be determined by imposing a new criterion, i.e., a multi-fluid algorithm should have the ability of maintaining a pure single-fluid. Based on these works, a new general algorithm, without any special treatment, for consistently implementing a high-order characteristic flux-split-based FDM is proposed in section 4.
\end{enumerate}

\section{Model}

\subsection{Physical model}

The multispecies model, expressed here in a one-dimensional form, is considered
\begin{align}
& \frac{\partial \rho }{\partial t} + \frac{\partial \rho u }{\partial x} = 0, \label{rho_eqn} \\
& \frac{\partial (\rho u)}{\partial t} + \frac{ \partial (\rho u^2 +p)}{\partial x} = 0, \label{rhou_eqn} \\
& \frac{\partial (\rho E)}{\partial t}  + \frac{ \partial (\rho Eu +pu) }{\partial x} =0, \label{rhoe_eqn} \\
& \frac{\partial (\rho Y_1)}{\partial t} + \frac{ \partial (\rho Y_1 u) }{\partial x} =0, \label{rhoy_eqn}
\end{align}
where $\rho$ is the density, $u$ is the velocity, $p$ is the pressure, $E = \rho e + \rho \frac{u^2}{2}$ is the total energy, and $e$ is the internal energy. Mass concentration $Y_k$ ($ \in [0,1]$) of the $k$th component satisfies $\sum_k Y_k =1$. To close this system, we require a closure between $p$ and $e$ of the mixture. This implies that a closure for the mixture is required.

In this study, we assume that there are $k(k=1,2)$ different fluid components. For each component $k$, the state variables such as density, pressure, internal energy, and temperature are denoted by $\rho_k$, $p_k$, $e_k$, and $e_k$, respectively. Furthermore, each component is assumed to satisfy the ideal gas law, i.e.,
\begin{align}
& \rho_k e_k = \frac{p_k}{\gamma_k -1} = C_{V,k}T_k,
\end{align}
where $C_{V,k}$ is the specific heat at constant volume and $\gamma_k$ is the ratio of the specific heats of the $k$th component gas.

The multispecies model is based on the local thermodynamic equilibrium. The fluids are initially well mixed, and the temperature is rapidly homogenized because of the occurrence of numerous collisions between various molecules \cite{Saurel1999JCP}. This yields $T_k =T$. Furthermore, mixture pressure $p$ is simply the sum of the partial pressures $p_k$ obtained using Dalton's law. These mixing rules finally yield the following relations:
\begin{align}
& \rho e = \left\{
             \begin{array}{ll}
        \frac{1}{\gamma -1} p = \left( \frac{\sum_k \frac{Y_k}{\gamma_k-1} \frac{1}{W_k}}{\sum_k \frac{Y_k}{W_k}} \right) p, & \hbox{Pressure-wise;} \\
               C_{V}T = \left( \sum_k Y_k C_{V,k} \right) T, & \hbox{Temparature-wise,}
             \end{array}
           \right.
   \label{EOS}
\end{align}
where $W_k$ is the molecular mass of $k$th component of the gas, and the mixture molecular mass $W$ follows from the definition of the mass fraction
\begin{align}
\frac{1}{W} =\sum_k \frac{Y_k}{W_k}.
\end{align}

Eqs.(\ref{rho_eqn}), (\ref{rhou_eqn}), (\ref{rhoe_eqn}), and (\ref{rhoy_eqn}) along with the appropriate relations between the mass concentration and material properties in the equation of state (Eq. \ref{EOS}) form a closed system.

\subsection{Numerical models}

It is well known that the discretization of the above system may result in spurious pressure oscillations \cite{Abgrall1996} and spurious temperature oscillations that may cause problems when the physical diffusion (i.e., for the Navier--Stokes equations) is included \cite{Johnsen2012}.

By numerical discretizing a material discontinuity problem in which the velocity and pressure are constant across the discontinuity, Abgrall \cite{Abgrall1996} was the first to recognize that the following transport equation
\begin{align}
& \frac{\partial}{\partial t} \left( \frac{1}{\gamma-1} \right) + u \frac{\partial}{\partial x} \left( \frac{1}{\gamma-1} \right) =0, \label{gamma_eqn}
\end{align}
should be solved to prevent the pressure oscillations. Shyue \cite{Shyue1998} later extended this idea to solving the transport equation for the mass fraction in advection form as well, and to complicated materials \cite{Shyue2001}. In order to further prevent the temperature oscillations, Johnsen and Ham \cite{Johnsen2012} proposed an overestimation formulation: solving the Euler equations along with two transport equations, one in conservative form for $\rho Y_1$ (i.e., Eq. (\ref{rhoy_eqn})), which is used to compute the temperature and mass fraction in the diffusive terms only, and one in advection form for $\frac{1}{\gamma - 1}$ (i.e., Eq. (\ref{gamma_eqn})), which is used to compute the pressure in the convective terms only. Using the same trick introduced by Abgrall, Beig and Johnsen \cite{Johnsen2015JCP} systematically analyzed the temperature oscillations phenomenon, and extended Abgrall's approach to two  supplementary transport equations, one in advection form for $\frac{1}{\gamma - 1}$ (i.e., Eq. (\ref{gamma_eqn})), and one is
\begin{align}
& \frac{\partial \left( \rho C_{V} \right)}{\partial t} + \frac{\partial \left( \rho C_{V} u \right) }{\partial x}  =0. \label{cv_eqn}
\end{align}

\begin{figure}[!ht]
  \centering
  \includegraphics[width=0.9\textwidth]{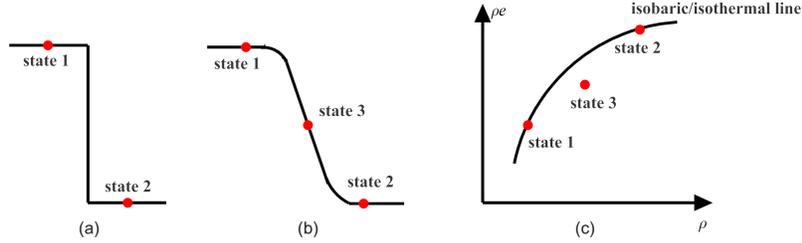}
  \caption{Schematic of the numerical errors due to the nonlinear EOS and numerical viscosity.} \label{fig1}
\end{figure}

In fact, we think that a better understanding of the numerical errors generated in compressible multi-material calculations using shock-capturing schemes is as follows. The mixture closure unusually leads to a highly nonlinear equation of state (see Eq. \ref{EOS}). This results in a problem that is also encountered in a single-material flow with a nonlinear equation of state \cite{hezw2016}. Let us consider the case in which there is an initial sharp interface (see Fig. \ref{fig1}(a)). If the classical fully conservative shock-capturing method is used, its numerical approximation will result in a numerical transition zone (see Fig. \ref{fig1}(b)). Generally, a new state in the numerical transition zone does not necessarily lie on the isobaric and/or isothermal line in $\rho-\rho e$ plane. Therefore, pressure and temperature spikes will be generated. These errors in turn will affect the internal energy, kinetic energy, and subsequently, the velocity.

From Eqs.(\ref{gamma_eqn}-\ref{cv_eqn}), we can see that the nonlinear relations between the internal energy, pressure, and/or temperature caused by the mixture closure (Eq. \ref{EOS}) are replaced by solving suitable transport equations in conservative or non-conservative forms. This is the key concept for avoiding the pressure and temperature oscillations. Furthermore, when the multi-material Navier--Stokes equations \cite{Tritschler2014, Cook2009} are solved, we require the mass fraction to obtain the mixture viscosity, mixture thermal conductivity, and effective binary diffusion coefficients. Therefore, the conservative equation of mass fraction should not be removed. In this paper, we directly solve the following model: Eqs.(\ref{rho_eqn}), (\ref{rhou_eqn}), (\ref{rhoe_eqn}), (\ref{rhoy_eqn}), (\ref{gamma_eqn}) and (\ref{cv_eqn}) form the final system, and are termed as the augmented multispecies model for convenience. In this system: (I) the relation between the pressure and energy is calculated only through the equation for $\frac{1}{\gamma-1}$ and (II) the the temperature is obtained only through the equation for $\rho C_{V}$. Finally, this model can be expressed as
\begin{align}
\frac{\partial \textbf{u}}{\partial t} + \textbf{A} \frac{\partial \textbf{u}}{\partial x} =0,
\end{align}
where
\begin{align}
\textbf{u}=\left(
             \begin{array}{c}
               \rho \\
               \rho u \\
               \rho E \\
               \rho Y_1 \\
               \frac{1}{\gamma-1} \\
               \rho C_V \\
             \end{array}
           \right),
\end{align}
and
\begin{align}
\textbf{A} = \left(
                                                                            \begin{array}{cccccc}
                                                                              0 & 1 & 0 & 0 & 0 & 0 \\
                                                                              -\frac{u^2}{2}(3-\gamma) & u(3-\gamma) & \gamma-1 & 0 & -(\gamma-1)p & 0 \\
                                                                              \frac{\gamma-1}{2}u^3 -uH & H-(\gamma -1)u^2 & \gamma u & 0 & -(\gamma -1)pu & 0 \\
                                                                              -Y_1 u & Y_1 & 0 & u & 0 & 0 \\
                                                                              0 & 0 & 0 & 0 & u & 0 \\
                                                                              -C_V u & C_V & 0 & 0 & 0 & u \\
                                                                            \end{array}
                                                                          \right).
\end{align}
After some manipulations, the eigen system of this model can be obtained as follows:
\begin{align}
\textbf{R} = [\textbf{r}_1, \textbf{r}_2, \textbf{r}_3, \textbf{r}_4, \textbf{r}_5, \textbf{r}_6] = \left(
                                                \begin{array}{cccccc}
                                                  1 & 1 & 1 & 0 & 0 & 0 \\
                                                  u-c & u & u+c & 0 & 0 & 0 \\
                                                  H-uc & \frac{u^2}{2} & H+uc & 0 & p & 0 \\
                                                  Y_1 & 0 & Y_1 & 1 & 0 & 0 \\
                                                  0 & 0 & 0 & 0 & 1 & 0 \\
                                                  C_V & 0 & C_V & 0 & 0 & 1 \\
                                                \end{array}
                                              \right),
\end{align}
\begin{align}
\Lambda = \left(
                                                \begin{array}{cccccc}
                                                  u-c &   &   &   &   &   \\
                                                    & u &   &   &   &   \\
                                                    &   &  u+c &   &   &   \\
                                                    &   &   &  u &   &   \\
                                                    &   &   &   &  u &   \\
                                                    &   &   &   &   &  u \\
                                                \end{array}
                                              \right),
\end{align}
and
\begin{align}
\textbf{L} = \left(
               \begin{array}{c}
                 \textbf{l}_1 \\
                 \textbf{l}_2 \\
                 \textbf{l}_3 \\
                 \textbf{l}_4 \\
                 \textbf{l}_5 \\
                 \textbf{l}_6 \\
               \end{array}
             \right)
 = \left(
                                                \begin{array}{cccccc}
                                                  \frac{u}{2c}+\chi \frac{u^2}{2} & -\frac{1}{2c} -\chi u & \chi & 0 & -\chi p & 0 \\
                                                  1-\chi u^2 & 2 \chi u & -2 \chi & 0 & 2 \chi p & 0 \\
                                                  -\frac{u}{2c}+\chi \frac{u^2}{2} & \frac{1}{2c} -\chi u & \chi & 0 & -\chi p & 0 \\
                                                  -\chi u^2 Y_1 & 2 \chi u Y_1 & -2 \chi u Y_1 & 1 & 2 \chi p Y_1 & 0 \\
                                                  0 & 0 & 0 & 0 & 1 & 0 \\
                                                  -\chi u^2 C_V & 2 \chi u C_V & -2 \chi u C_V & 0 & 2 \chi p C_V & 1 \\
                                                \end{array}
                                              \right),
\end{align}
where $H = h+\frac{u^2}{2} = (e + p/\rho) + \frac{u^2}{2}$, $\chi= \frac{\gamma-1}{2c^2}$, and $c$ is the speed of sound.

\textbf{Remark}. Here, we should point out that there are still some other sources of numerical oscillations such as the start-up error \cite{Johnsen2008, Nonomura2012}, component-wise nonlinear discretization error \cite{Hezw2015}, and nonlinear conservative variables reconstruction error \cite{Johnsen2011}. These errors do affect the velocity, pressure, and temperature equilibria to some extent. However, these errors can get generated even in single-material ideal gas flows. Therefore, these errors are not purely related to the mixture closure in compressible multi-material gas flows.

\section{Review of previous consistent characteristic finite difference algorithms}
All the numerical models discussed above are non-conservative or quasi-conservative. Therefore, it is not an easy work to design a consistent flux-split based finite difference algorithm, because there are no Riemann solver and primitive-variable reconstruction in this kind of numerical methods. Till date, there are only two attempts on consistent implementing the characteristic flux-split based finite difference to solve these numerical models, while special treatments are needed \cite{hezw2016, Nonomura2017}. In this section, we report a detailed analysis of these two algorithms, to reveal the reason for the special treatments in these methods.

For clarity, we first describe the numerical framework used in this paper. The spatial domain was discretized into an $N$-point grid with uniform grid spacing $\Delta x$, so that the augmented multispecies model at point $x_i$ could be updated in one time step ($\Delta t$) or a substep.
\begin{align}
\textbf{u}_{i}^{n+1} = \textbf{u}_{i}^{n} - \Delta t \frac{\textbf{F}_i}{\Delta x} ,
\end{align}
where
\begin{align}
\textbf{F}_i= \left(
                \begin{array}{c}
                  F_i^{\rho} \\
                  F_i^{\rho u} \\
                  F_i^{\rho E} \\
                  F_i^{\rho Y_1} \\
                  F_i^{\frac{1}{\gamma-1}} \\
                  F_i^{\rho C_V} \\
                \end{array}
              \right)
              = \left(
                \begin{array}{c}
                  \widehat{f}^{\rho}_{i+1/2} - \widehat{f}^{\rho}_{i-1/2} \\
                 \widehat{f}^{\rho u}_{i+1/2} - \widehat{f}^{\rho u}_{i-1/2} \\
                  \widehat{f}^{\rho E}_{i+1/2} - \widehat{f}^{\rho E}_{i-1/2} \\
                  \widehat{f}^{\rho Y_1}_{i+1/2} - \widehat{f}^{\rho Y_1}_{i-1/2} \\
                  F_i^{\frac{1}{\gamma-1}} \\
                  F_i^{\rho C_V} \\
                \end{array}
              \right).
\end{align}

We mainly use the characteristic flux-split based FDM \cite{Shu1996, hezw2016}. However, the equation for $\frac{1}{\gamma-1}$ is non-conservative. Therefore, the specific discretization form of $\textbf{F}_i$ is unknown. One approach to obtain the consistent discretization form of $\textbf{F}_i$ is to use Abgrall's criterion \cite{Abgrall1996}, i.e., the velocity and pressure equilibria should be maintained during time updates. Therefore, in the following section, we mainly focus on determining the specific expression for $\textbf{F}_i$ so that if $u_i^n=u_0$, $p_i^n=p_0$, $T_i^n=T_0$, then $u_i^{n+1}=u_0$, $p_i^{n+1}=p_0$, $T_i^{n+1}=T_0$.

\subsection{Method of He et al. \cite{hezw2016}} \label{hezw_method}

According to this method, the augmented model should be written as
\begin{align}
\frac{\partial \textbf{u}}{\partial t} + \frac{\partial \textbf{f}(\textbf{u})}{\partial x} + \textbf{B} \frac{\partial \textbf{u}}{\partial x} =0,
\end{align}
where
\begin{align}
 \textbf{f}=\left(
             \begin{array}{c}
               \rho u \\
               \rho u^2+p \\
               \rho Eu+pu \\
               \rho Y_1 u \\
               0 \\
               \rho C_V u \\
             \end{array}
           \right),
 \textbf{B}=\left(
              \begin{array}{cccccc}
                0 &  &  &  &  &  \\
                 & 0 &  &  &  &  \\
                 &  & 0 &  &  &  \\
                 &  &  & 0 &  &  \\
                 &  &  &  & u &  \\
                 &  &  &  &  & 0 \\
              \end{array}
            \right).
\end{align}

Using the finite-difference methodology with a local characteristic decomposition, we can obtain the positive/negative numerical flux  $\widehat{\textbf{f}}_{i+1/2}^{\pm}$ as following
\begin{align}
\widehat{\textbf{f}}_{i+1/2}^{\pm} = \sum_{s=1}^6 \widehat{q}_{s,i+1/2}^{\pm} \textbf{r}_{s,i+1/2} = \left(
                                       \begin{array}{c}
                                         \sum_{s=1}^3 \widehat{q}_{s,i+1/2}^{\pm} \\
                                         u_0 \sum_{s=1}^3 \widehat{q}_{s,i+1/2}^{\pm} + c_{i+1/2} Q_{M,i+1/2}^{\pm}  \\
                                         \frac{u_0^2}{2} \sum_{s=1}^3 \widehat{q}_{s,i+1/2}^{\pm} + u_0 c_{i+1/2} Q_{M,i+1/2}^{\pm} + h_{i+1/2} Q_{P,i+1/2}^{\pm} + p_0 \widehat{q}_{5,i+1/2}^{\pm}  \\
                                         Y_{1,i+1/2} Q_{P,i+1/2}^{\pm} + \widehat{q}_{4,i+1/2}^{\pm} \\
                                         \widehat{q}_{5,i+1/2}^{\pm} \\
                                         C_{V,i+1/2} Q_{P,i+1/2}^{\pm} + \widehat{q}_{6,i+1/2}^{\pm} \\
                                       \end{array}
                                     \right),
\label{numerical_nf}
\end{align}
where
\begin{align}
& Q_{M,i+1/2}^{\pm} = \widehat{q}_{3,i+1/2}^{\pm} - \widehat{q}_{1,i+1/2}^{\pm}, \\
& Q_{P,i+1/2}^{\pm} = \widehat{q}_{3,i+1/2}^{\pm} + \widehat{q}_{1,i+1/2}^{\pm},
\end{align}
and $\widehat{q}_{s,i+1/2}^{\pm} (s=1, \cdots, 6)$ is the characteristic positive/negative numerical flux in each characteristic field. The characteristic positive/negative numerical flux $\widehat{q}_{s,i+1/2}^{\pm}$ in the $s$th characteristic field is reconstructed using any nonlinear scheme $D_{s,i+1/2}^{\pm}[\cdot]$ (such as the WENO \cite{Shu1996} and monotonicity-preserving (MP-R) \cite{hezw_mp2} schemes) with the following characteristic variable:
\begin{align}
q^{\pm}_{s,i+l}=\frac{1}{2} \textbf{l}_{s,i+1/2} \cdot \left( \textbf{f}_{i+l} \pm \alpha_{s,i+1/2} \textbf{u}_{i+l}\right),
\label{chara_var}
\end{align}
where variables at $i+1/2$ are calculated by the Roe-average or simple mean of those on the nearest two points, $j$ and $j+1$. Here,
\begin{align}
\alpha_{s,i+1/2} =\kappa \times \hbox{max}( \left|\lambda_{s,i-3} \right|, \cdots,  \left|\lambda_{s,i+2} \right|),
\end{align}
where $\lambda_s$ is the $s$th eigenvalue for the $s$th characteristic wave, and $\kappa$ is a constant with value typically from 0.9 to 1.1.

Under the condition that $u=u_0$, $p=p_0$, the characteristic variable vector can be simplified as
\begin{align}
\textbf{q}_{i+l}^{\pm} = \left(
                           \begin{array}{c}
                             q^{\pm}_{1,i+l} \\
                             q^{\pm}_{2,i+l} \\
                             q^{\pm}_{3,i+l} \\
                             q^{\pm}_{4,i+l} \\
                             q^{\pm}_{5,i+l} \\
                             q^{\pm}_{6,i+l} \\
                           \end{array}
                         \right)
 = \frac{1}{2} \left(
                                       \begin{array}{c}
                                         -\frac{p_0}{2c_{i+1/2}} + \chi_{i+1/2} u_0 \left(\rho e \right)_{i+l} \\
                                         \left( u_0  \pm \alpha_{2,i+1/2} \right) \rho_{i+l}  - 2 \chi_{i+1/2} u_0 \left(\rho e \right)_{i+l} \\
                                         \frac{p_0}{2c_{i+1/2}} + \chi_{i+1/2} u_0 \left(\rho e \right)_{i+l} \\
                                         \left( u_0  \pm \alpha_{4,i+1/2} \right) (\rho Y_1)_{i+l}  - 2 \chi_{i+1/2} Y_{1,i+1/2} u_0 \left(\rho e \right)_{i+l} \\
                                         \left( 0  \pm \alpha_{5,i+1/2} \right) \left( \frac{1}{\gamma-1} \right)_{i+l} \\
                                         \left( u_0  \pm \alpha_{6,i+1/2} \right) (\rho C_V)_{i+l} - 2 \chi_{i+1/2} C_{V,i+1/2} u_0 \left(\rho e \right)_{i+l} \\
                                       \end{array}
                                     \right).
\label{hoho}
\end{align}

It was pointed out in \cite{hezw2016}, $Q_{M,i+1/2}^{\pm}$ and $Q_{P,i+1/2}^{\pm}$ can be simplified if the schemes in genuinely nonlinear characteristic fields are the same, i.e.,
\begin{align}
D_{1,i+1/2}^{\pm}[\cdot] = D_{3,i+1/2}^{\pm}[\cdot] \equiv D_{0,i+1/2}^{\pm}[\cdot].
\end{align}
Under this constraint, we can obtain
\begin{align}
& Q_{M,i+1/2}^{\pm}= \frac{p_0}{c_{i+1/2}}  \label{sim1} \\
& Q_{P,i+1/2}^{\pm}= 2 \chi_{i+1/2} u_0 D_{0,i+1/2}^{\pm}[ \left(\rho e \right)_{i+l}]. \label{sim2}
\end{align}
Substituting Eqs.(\ref{sim1}) and (\ref{sim2}) into Eq. (\ref{numerical_nf}), we obtain that $F_{i}^{\rho u} = u_0 F_{i}^{\rho}$. Therefore, the velocity equilibrium is maintained.

Under the velocity equilibrium, we can further derive
\begin{align}
F_i^{\rho E} = & \frac{u_0^2}{2} F_i^{\rho} + h_{i+1/2} \left( Q_{P,i+1/2}^{+} + Q_{P,i+1/2}^{-} \right) - h_{i-1/2} \left( Q_{P,i-1/2}^{+} + Q_{P,i-1/2}^{-} \right) \nonumber \\
               & + p_0 \left( \widehat{q}_{5,i+1/2}^{+} +\widehat{q}_{5,i+1/2}^{-} -\widehat{q}_{5,i-1/2}^{+} - \widehat{q}_{5,i-1/2}^{-} \right).
\end{align}
Therefore, we have
\begin{align}
(\rho e)_i^{n+1} =  (\rho e)_i^{n} & -\frac{\Delta t}{\Delta x} \left( h_{i+1/2} \left( Q_{P,i+1/2}^{+} + Q_{P,i+1/2}^{-} \right) - h_{i-1/2} \left( Q_{P,i-1/2}^{+} + Q_{P,i-1/2}^{-} \right)\right) \nonumber \\
                                   & -  \frac{\Delta t}{\Delta x} p_0 \left( \widehat{q}_{5,i+1/2}^{+} +\widehat{q}_{5,i+1/2}^{-} -\widehat{q}_{5,i-1/2}^{+} - \widehat{q}_{5,i-1/2}^{-} \right).
\label{innerenergy}
\end{align}
To maintain the pressure equilibrium, Eq. (\ref{innerenergy}) should hold for an arbitrary pressure. By extracting the pressure from Eq. (\ref{innerenergy}), the proper discretization form for the equation for $\frac{1}{\gamma-1}$ is as follows:
\begin{align}
F_i^{\frac{1}{\gamma-1}} =  u \left( \frac{D_{0,i+1/2}^{+} \left[ \left( \frac{1}{\gamma-1} \right)_{i+l} \right] + D_{0,i+1/2}^{-} \left[ \left( \frac{1}{\gamma-1} \right)_{i+l} \right]}{2} - \frac{D_{0,i-1/2}^{+} \left[ \left( \frac{1}{\gamma-1} \right)_{i+l} \right] + D_{0,i-1/2}^{-} \left[ \left( \frac{1}{\gamma-1} \right)_{i+l} \right]}{2} \right) \nonumber \\
 + \left(\alpha_{5,i+1/2} \frac{D_{i+1/2}^{+} \left[ \left( \frac{1}{\gamma-1} \right)_{i+l} \right] - D_{i+1/2}^{-} \left[ \left( \frac{1}{\gamma-1} \right)_{i+l} \right]}{2} - \alpha_{5,i-1/2} \frac{D_{i-1/2}^{+} \left[ \left( \frac{1}{\gamma-1} \right)_{i+l} \right] - D_{i-1/2}^{-} \left[ \left( \frac{1}{\gamma-1} \right)_{i+l} \right]}{2} \right).
\end{align}

These results reconfirm the conclusion that the quasi-conservative approach can be realized using characteristic-wise-based FDM if the schemes in the genuinely nonlinear characteristic fields can be ensured to be same and the decomposition equation representing the material interfaces is appropriately discretized \cite{hezw2016}. Under the velocity and pressure equilibria, we can further explore the constraint for maintaining the equilibrium of temperature. Similar to the derivation of the pressure equilibrium, we are only required to extract the arbitrary temperature from Eq. (\ref{innerenergy}). After this operation, we can obtain the consistent discretization of $F_i^{\rho C_V}$.

\subsection{Method of Nonomura et al. \cite{Nonomura2017}}

According to this method, the augmented model should be written as
\begin{align}
\frac{\partial \textbf{u}}{\partial t} + \textbf{M} \frac{\partial \textbf{f}(\textbf{u})}{\partial x} =0,
\label{nono_eqn}
\end{align}
where
\begin{align}
\textbf{u}=\left(
             \begin{array}{c}
               \rho \\
               \rho u \\
               \rho E \\
               \rho Y_1 \\
               \frac{1}{\gamma-1} \\
               \rho C_V \\
             \end{array}
           \right),
 \textbf{f}=\left(
             \begin{array}{c}
               \rho u \\
               \rho u^2+p \\
               \rho Eu+pu \\
               \rho Y_1 u \\
               \frac{1}{\gamma-1} \\
               \rho C_V u \\
             \end{array}
           \right),
 \textbf{M}=\left(
              \begin{array}{cccccc}
                1 &  &  &  &  &  \\
                 & 1 &  &  &  &  \\
                 &  & 1 &  &  &  \\
                 &  &  & 1 &  &  \\
                 &  &  &  & u &  \\
                 &  &  &  &  & 1 \\
              \end{array}
            \right).
\end{align}

By trial and error, Nonomura and Fujii \cite{Nonomura2017} proposed the following discretization form
\begin{align}\label{nono_method}
  \textbf{M}_i \left( \frac{\partial \textbf{f}}{\partial x} \right)_i = \textbf{M}_i^n \frac{\textbf{F}_{i+1/2}^C -\textbf{F}_{i-1/2}^C}{\Delta x} + \frac{\textbf{F}_{i+1/2}^D -\textbf{F}_{i-1/2}^D}{\Delta x},
\end{align}
where $\textbf{F}_{i \pm 1/2}^C$ is a linear combination (sixth-order central linear scheme in \cite{Nonomura2017}) of the original flux $\textbf{f}_i^n$ in Eq. (\ref{nono_eqn}), and $\textbf{F}_{i \pm 1/2}^D$, which is only used for numerical dissipation, is constructed by using the difference between the nonlinear WENO scheme and sixth-order central linear scheme and by employing a modified conservative flux on the node as follows:
\begin{align}
  q_{s,i+l}^{\pm} =\frac{1}{2} \textbf{l}_{s,i+1/2} \cdot \left( \textbf{F}_{i+l,i \pm 1/2}^{m} \pm \textbf{u}_{i+l}^{m} \right),
\end{align}
\begin{align}
  \textbf{F}_{i+l, i \pm 1/2}^{m} = \left(
                                      \begin{array}{c}
                                        \rho_{i+l} \\
                                        \rho_{i+l}u_{i+l}^{2} +p_{i+l} \\
                                        u_{i+l}(\rho_{i+l} E_{i+l} +p_{i+l}) \\
                                        \rho_{i+l} u_{i+l} Y_{1,i+l} \\
                                        \widetilde{u}_{i \pm 1/2} \left( \frac{1}{\gamma -1} \right)_{i+l} \\
                                        \rho_{i+l}C_{V,i+l} u_{i+l}\\
                                      \end{array}
                                    \right),
\end{align}
where $\widetilde{u}_{i \pm 1/2}$ is the Roe averaged velocity \cite{Nonomura2017}.

Similar to the derivation process in the previous method, the authors prove that this discretization form can maintain the velocity, pressure, and temperature equilibria within the overestimated quasi-conservative formulation \cite{Johnsen2012}, though some special treatments are also required \cite{Nonomura2017}.

\subsection{Brief remark}

Here, we provide a brief remark on the above-mentioned two methods. First, it is apparent that the simplification of $Q_{M,i+1/2}^{\pm}$ and $Q_{P,i+1/2}^{\pm}$ plays a fundamental role in the first method. The zero flux in the $\frac{1}{\gamma-1}$ equation causes $Q_{M,i+1/2}^{\pm}$ and $Q_{P,i+1/2}^{\pm}$ to be only simplified under a strict constraint, i.e., the schemes in genuinely nonlinear characteristic fields must be ensured to be same. This constraint finally affects the discretization of $\frac{1}{\gamma-1}$ and $\rho C_V$ equations. Therefore, it can be expected that if the convection $\left( u \frac{1}{\gamma-1} \right)_x$ of $\frac{1}{\gamma-1}$ appears in the flux, then this constraint may be abandoned (see next section). Next, the use of the split form of the WENO scheme in the second method is simply to adopt the idea proposed by Terashima et al. \cite{Terashima2013} who directly implemented central finite difference schemes by introducing consistent local artificial diffusion terms to suppress the numerical oscillations of the pressure and velocity. The novel conclusion of the work \cite{Nonomura2017} is that they find that the consistent vector form of the numerical dissipation is closely related to the interpolation in the characteristic fields. In addition, here we should point out another treatment in the second method. In dissipation flux $\textbf{F}_{i \pm 1/2}^D$, the authors utilize constant $\widetilde{u}_{i\pm 1/2}$ in the entire stencil of the WENO scheme. This treatment numerically demonstrated that their final algorithm had sufficient and approximate numerical dissipation for problems with interactions of interfaces and strong shock and rarefaction waves.

Based on this analysis, a new consistent algorithm, without any special treatment, can be proposed (see next section).

\section{New consistent characteristic finite difference algorithm}

The main characteristics of this new consistent algorithm are: (I) to introduce the convection $\left( u \frac{1}{\gamma-1} \right)_x$ of $\frac{1}{\gamma-1}$ with constant $\widehat{u}_{i\pm 1/2}$ in the entire stencil of any scheme and (II) to obtain the consistent discretization form of the source term containing the velocity divergence using a new additional criterion. The following are the details.

Fist, we rewrite the augmented model as
\begin{align}
\frac{\partial \textbf{u}}{\partial t} + \frac{\partial \textbf{f}(\textbf{u})}{\partial x} = \textbf{S} (\textbf{u}),
\end{align}
where
\begin{align}
 \textbf{f}=\left(
             \begin{array}{c}
               \rho u \\
               \rho u^2+p \\
               \rho Eu+pu \\
               \rho Y_1 u \\
               \frac{1}{\gamma -1}u \\
               \rho C_V u \\
             \end{array}
           \right),
 \textbf{S}=\left(
             \begin{array}{c}
               0 \\
               0 \\
               0 \\
               0 \\
               \frac{1}{\gamma-1} \frac{\partial u}{\partial x} \\
               0 \\
             \end{array}
           \right).
\end{align}

Second, $\textbf{F}_i$ is discretized directly as follows:
\begin{align}\label{nono_method}
 \textbf{F}_i = \left( \widehat{\textbf{f}}_{i+1/2} - \widehat{\textbf{f}}_{i-1/2} \right) - \textbf{S}_i^n,
\end{align}
where $\widehat{\textbf{f}}_{i \pm 1/2}$ is constructed by using the following conservative flux on the node.
\begin{align}
  \textbf{f}_{i+l, i \pm 1/2}= \left(
                                      \begin{array}{c}
                                        \rho_{i+l} \\
                                        \rho_{i+l}u_{i+l}^{2} +p_{i+l} \\
                                        u_{i+l}(\rho_{i+l} E_{i+l} +p_{i+l}) \\
                                        \rho_{i+l} u_{i+l} Y_{1,i+l} \\
                                        \widehat{u}_{i \pm 1/2} \left( \frac{1}{\gamma -1} \right)_{i+l} \\
                                        \rho_{i+l}C_{V,i+l} u_{i+l}\\
                                      \end{array}
                                    \right),
\end{align}
where $\widehat{u}_{i \pm 1/2}$ is the constant velocity in the entire stencil with  properties: (1) $\widehat{u}_{i+1/2} = U(u_{i-sl}, \cdots, u_{i+ sr})$, where $\{i-sl, \cdots, i+ sr \}$ is the entire stencil in the conservative scheme; and (2) $\widehat{u}_{i+1/2} = U(u, \cdots, u) = u$.

Before we provide the specific expressions of $\widehat{u}_{i \pm 1/2}$ and $\textbf{S}_i^n$, we show that this general form satisfies Abgrall's criterion. The derivation process is the same as that in the previous method discussed in section \ref{hezw_method} and \cite{hezw2016}. In this new framework, the corresponding characteristic variable $q^{\pm}_{s,i+l}=\frac{1}{2} \textbf{l}_{s,i+1/2} \cdot \left( \textbf{f}_{i+l} \pm \alpha_{s,i+1/2} \textbf{u}_{i+l}\right)$ can be simplified under the condition that $u=u_0$, $p=p_0$, and $T=T_0$.
\begin{align}
\textbf{q}_{i+l}^{\pm}  = \frac{1}{2} \left(
                                       \begin{array}{c}
                                         -\frac{p_0}{2c_{i+1/2}} \\
                                         \left( u_0  \pm \alpha_{2,i+1/2} \right) \rho_{i+l} \\
                                         \frac{p_0}{2c_{i+1/2}} \\
                                         \left( u_0  \pm \alpha_{4,i+1/2} \right) (\rho Y_1)_{i+l}  \\
                                         \left( u_0  \pm \alpha_{5,i+1/2} \right) \left( \frac{1}{\gamma-1} \right)_{i+l} \\
                                         \left( u_0  \pm \alpha_{6,i+1/2} \right) (\rho C_V)_{i+l}  \\
                                       \end{array}
                                     \right).
\label{hoho_new}
\end{align}
From Eq. (\ref{hoho_new}), we can see that $Q_{M,i+1/2}^{\pm}$ and $Q_{P,i+1/2}^{\pm}$ can be simplified without any constraint.
\begin{align}
& Q_{M,i+1/2}^{\pm}= \frac{p_0}{c_{i+1/2}}  \label{sim1_new} \\
& Q_{P,i+1/2}^{\pm}= 0. \label{sim2_new}
\end{align}
Substituting Eqs.(\ref{sim1_new}) and (\ref{sim2_new}) into Eq. (\ref{numerical_nf}), we can easily obtain $F_i^{\rho u} =u_0 F_i^{\rho}$. Therefore, the velocity equilibrium is maintained. Under this condition, we can further derive $F_i^{\rho E} =\frac{u_0^2 }{2} F_i^{\rho} + p_0 F_i^{\frac{1}{\gamma-1}}$ and $F_i^{\rho E} =\frac{u_0^2 }{2} F_i^{\rho} + T_0 F_i^{\rho C_V}$. Therefore, the pressure and temperature equilibria are also maintained.

Finally, we discuss the specific expressions of $\widehat{u}_{i \pm 1/2}$ and $\textbf{S}_i^n$. Abgrall's criterion cannot yield any consistent discretization form of $\textbf{S}_i^n$. In this paper, we propose a simple additional criterion: \emph{a multi-fluid algorithm should have the ability of maintaining a pure single-fluid}. For the case considered in this study, this implies that if $\frac{1}{\gamma-1}$ is a constant, it should not change during time updates. According to this criterion, we can easily obtain
\begin{align}\label{nono_method}
\textbf{S}_i^n = \left(
                   \begin{array}{c}
                     0 \\
                     0 \\
                     0 \\
                     0 \\
                     \left(\frac{1}{\gamma-1}\right)_{i}^n  \left( \widehat{u}_{i + 1/2} - \widehat{u}_{i - 1/2}\right) \\
                     0 \\
                   \end{array}
                 \right).
\end{align}
Furthermore, because the simple Lax--Friedrichs flux split is used in this work, $\widehat{u}_{i \pm 1/2}$ can be correspondingly obtained by
\begin{align}
 \widehat{u}_{i \pm 1/2} = D_{i \pm 1/2} ^{+} \left[ \frac{1}{2}(u_{i+l} + \alpha_{5,i \pm 1/2}) \right] + D_{i \pm 1/2} ^{-} \left[ \frac{1}{2}(u_{i+l} - \alpha_{5,i \pm 1/2}) \right].
\end{align}

This completes the new consistent algorithm. The derivation process clearly reveals that the new algorithm does not require any special treatment, and can be applied directly to any kind of conservative finite difference schemes.

\section{Numerical tests and discussions}

Several numerical tests in one- and two-dimensional spaces are performed. For the method in which a common weight technique for the WENO schemes in all genuinely nonlinear characteristic fields should be utilized \cite{hezw2016}, the design of a more stable common weight technique is still an open question. For the method in which the split form of the WENO scheme should be employed \cite{Nonomura2017}, our numerical experiments show that the results reported in the work \cite{Nonomura2017} show no obvious differences with ones obtained by our new proposed algorithm that does not require any special treatment. Because these reasons, we only report the results of the fully conservative multispecies model that is solved by following the general WENO methodology \cite{Shu1996} and of the augmented multispecies model that is solved by the newly proposed consistent algorithm (coupled with the fifth-order WENO scheme \cite{Shu1996}). In the following subsection, "FC-WENO5" refers to the fully conservative multispecies model solved by following the general WENO methodology \cite{Shu1996}. Specifically, the simple mean is used for the characteristic decomposition at the cell faces, whereas the Lax--Friedrichs formulation is applied for the numerical fluxes. For more details, please refer \cite{Shu1996}. The term "present" implies that we use the newly proposed consistent algorithm (coupled with the fifth-order WENO scheme \cite{Shu1996}) to solve the augmented multispecies model. For two-dimensional cases, the dimension-by-dimension technique is used. For all the computations, a third-order TVD Runge--Kutta scheme \cite{Gottlieb&Shu1998} is used for the time integration, and the CFL number is set to 0.5.

\subsection{Moving material interface problem}

The following moving material interface problem with a very large density jump is first examined:
\begin{align}
  (\rho,u, p, \gamma, W)=\left\{
                           \begin{array}{ll}
                             (40.0, 1.0, 1.0/1.4, 1.667, 40), & -0.25< x<0.25 \\
                             (1.0, 1.0, 1.0/1.4, 1.4, 1), & \hbox{otherwise.}
                           \end{array}
                         \right.
\end{align}
The computational domain of $-0.5 \leq x \leq 0.5$ is set with 100 grid points, and the periodic boundary conditions are applied at $x=\pm 0.5$. The grid spacing is uniformly distributed, and results of $t=1$ are discussed.

\begin{figure}[!ht]
  \centering
  \includegraphics[width=0.48\textwidth]{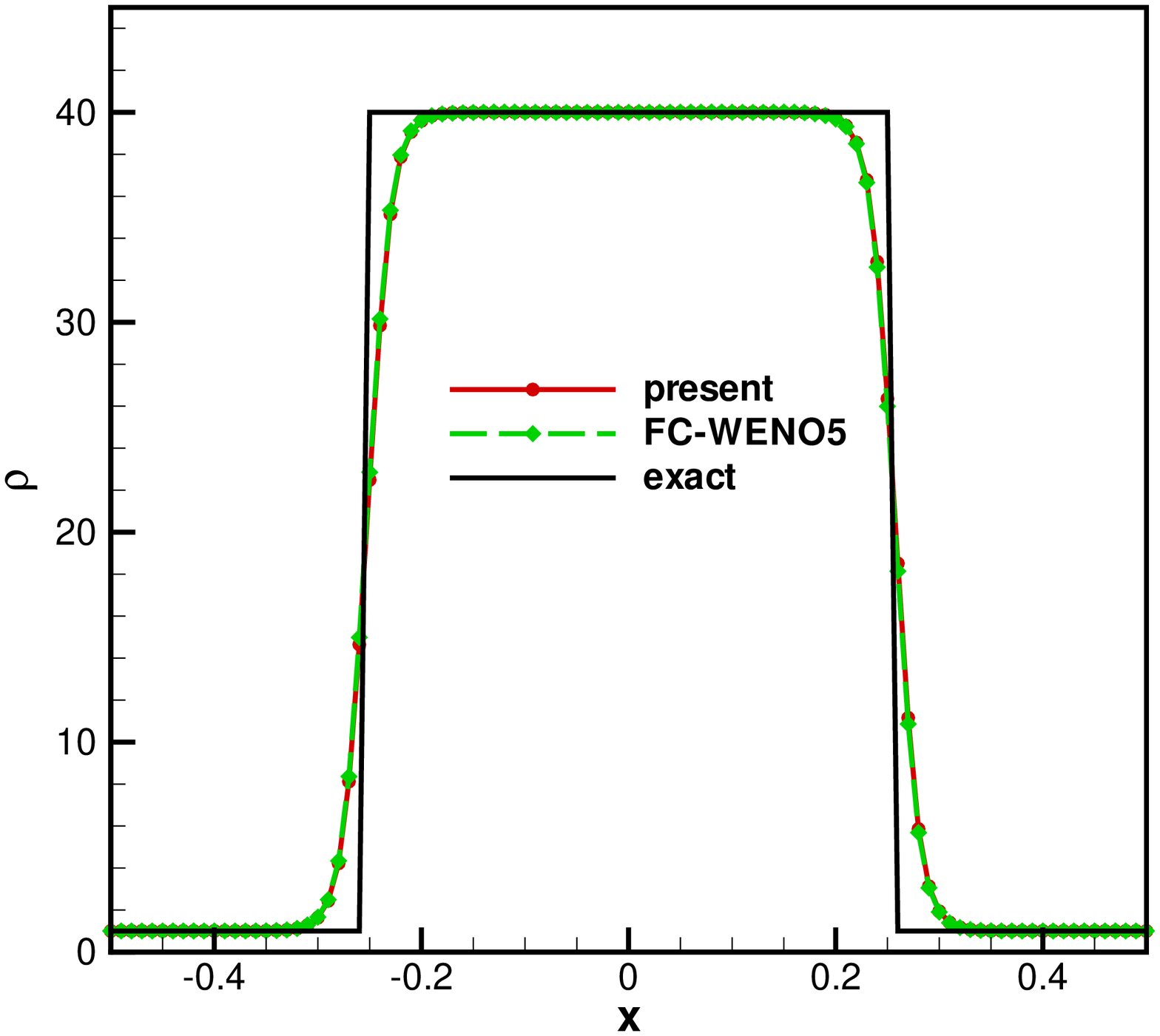}
  \includegraphics[width=0.48\textwidth]{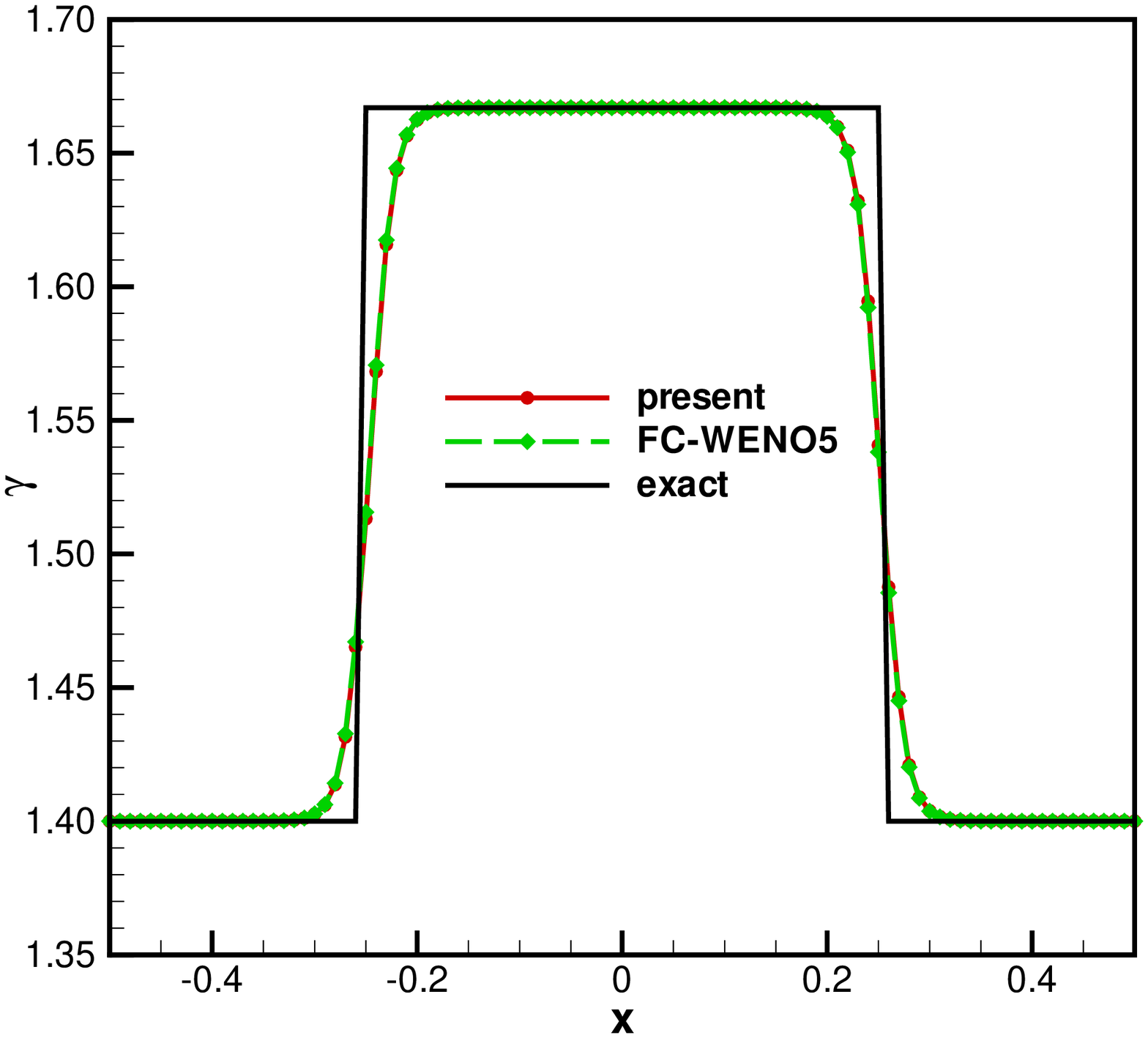}
  \includegraphics[width=0.48\textwidth]{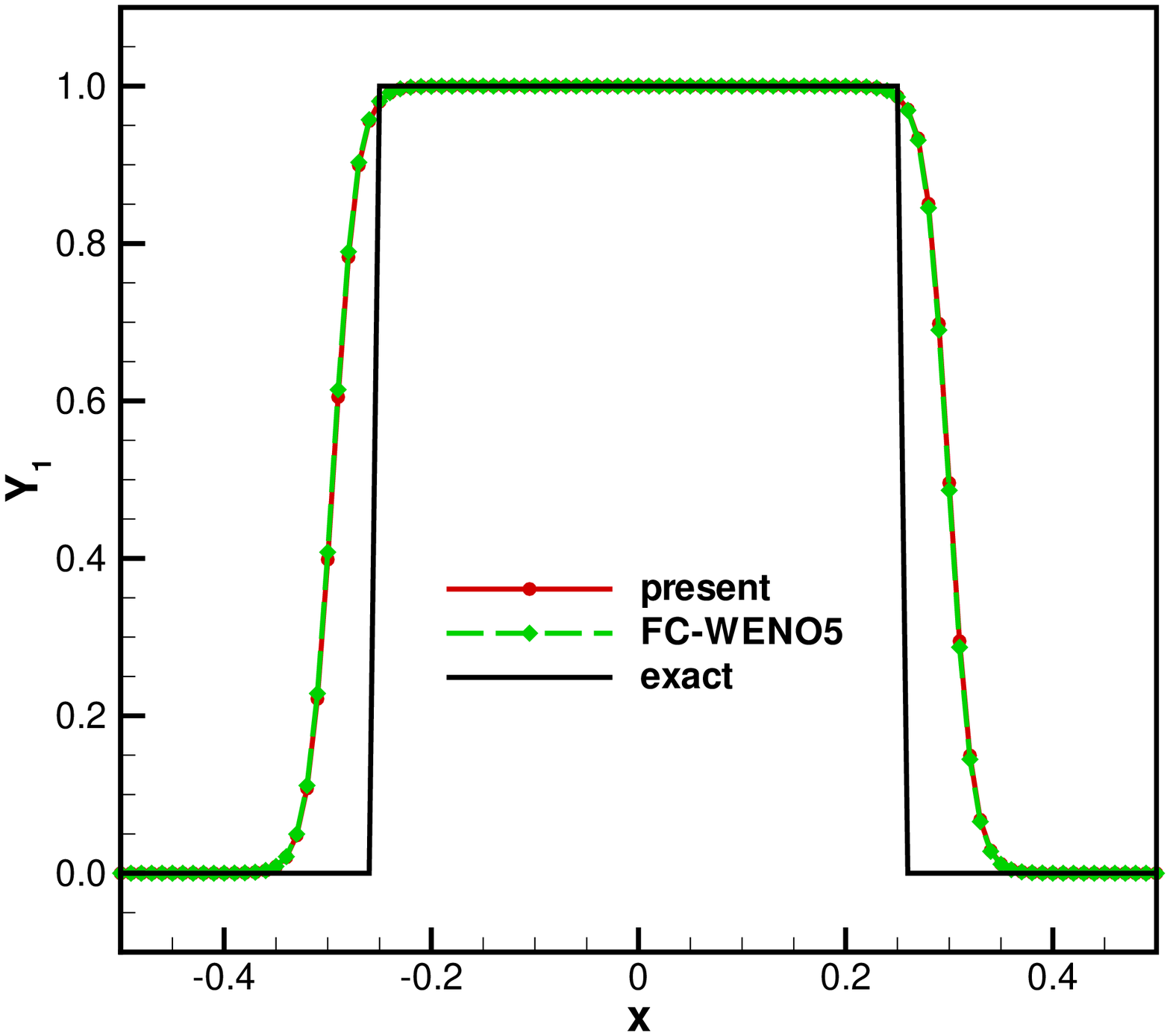}
  \includegraphics[width=0.48\textwidth]{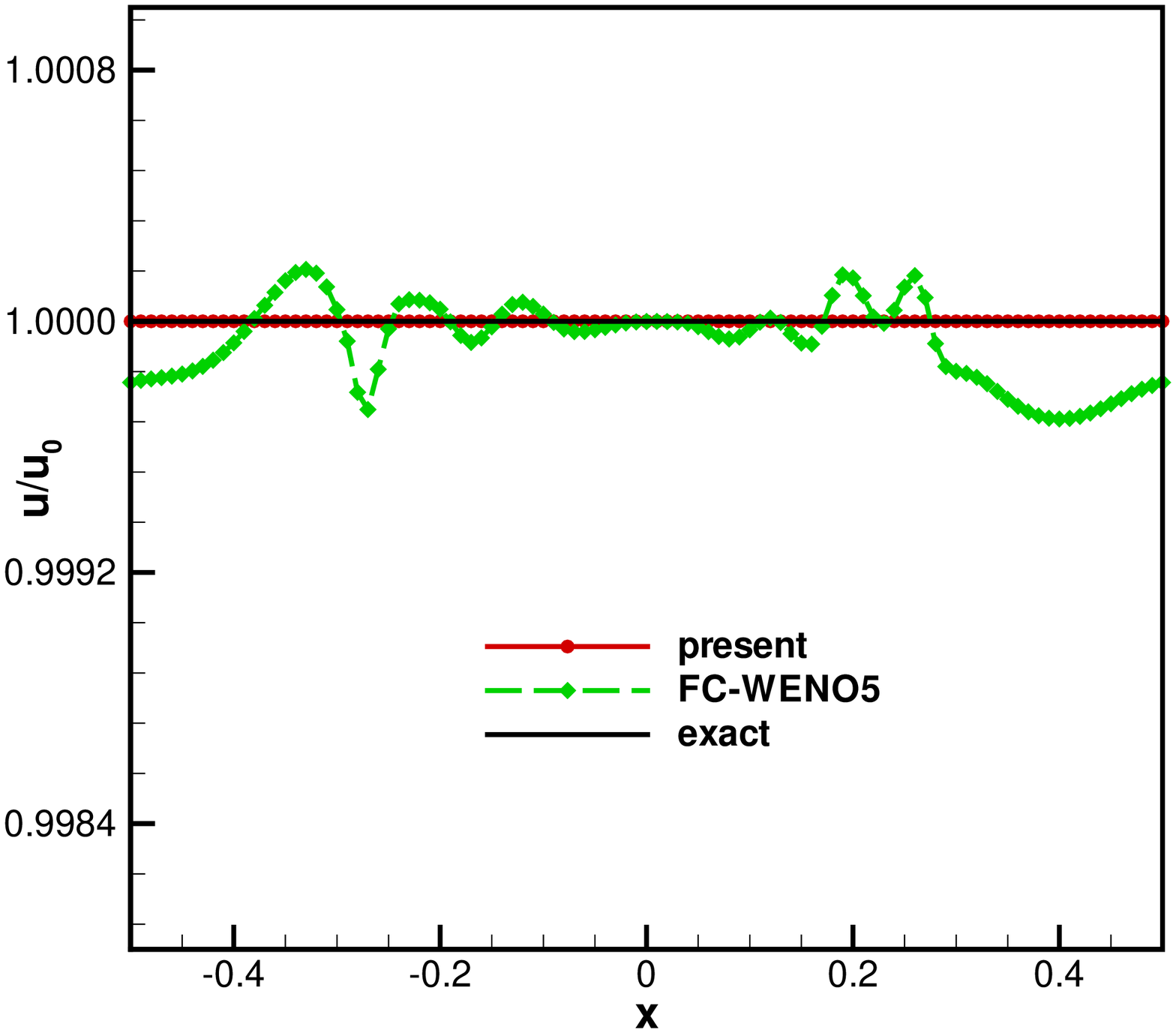}
  \includegraphics[width=0.48\textwidth]{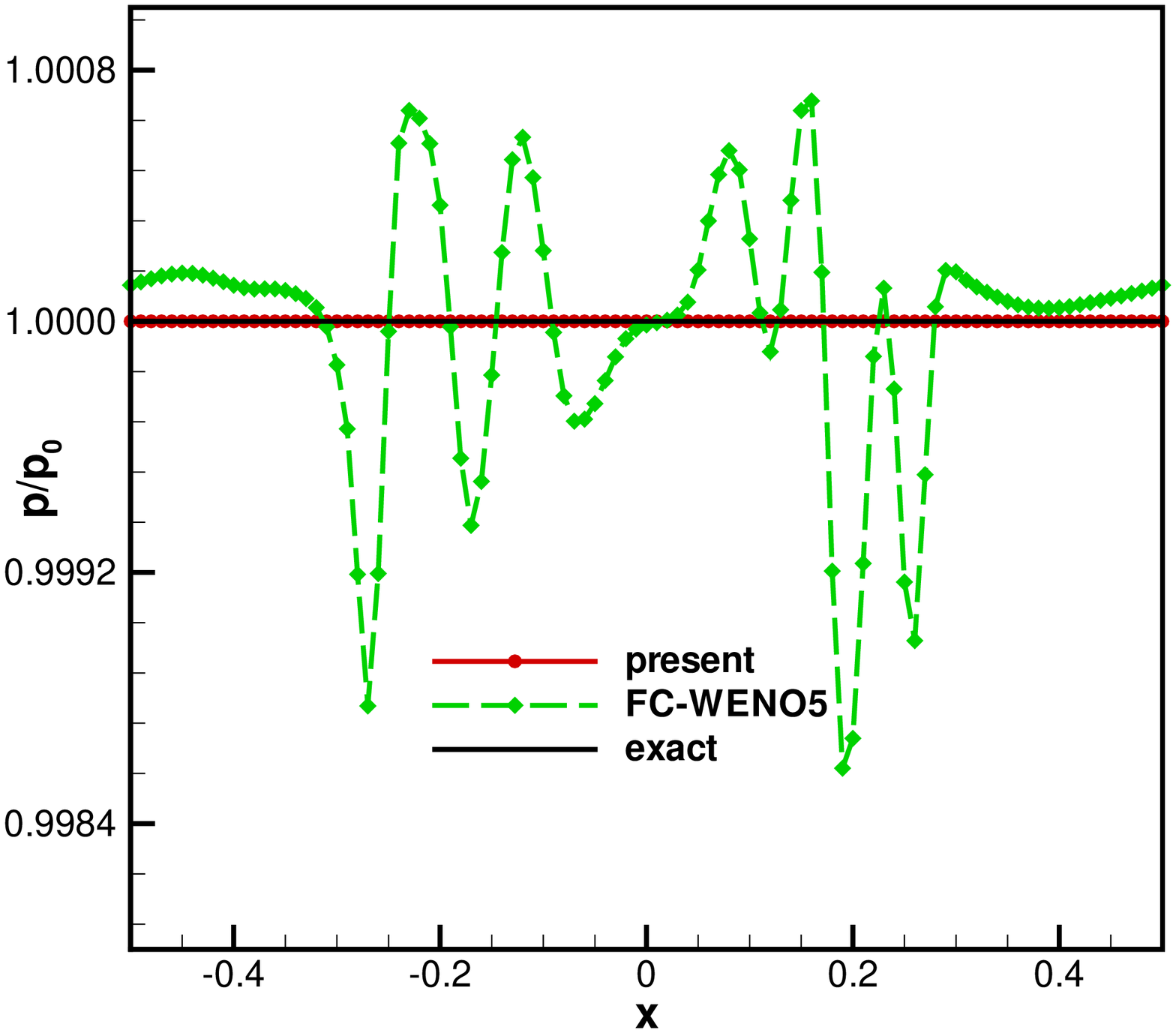}
  \includegraphics[width=0.48\textwidth]{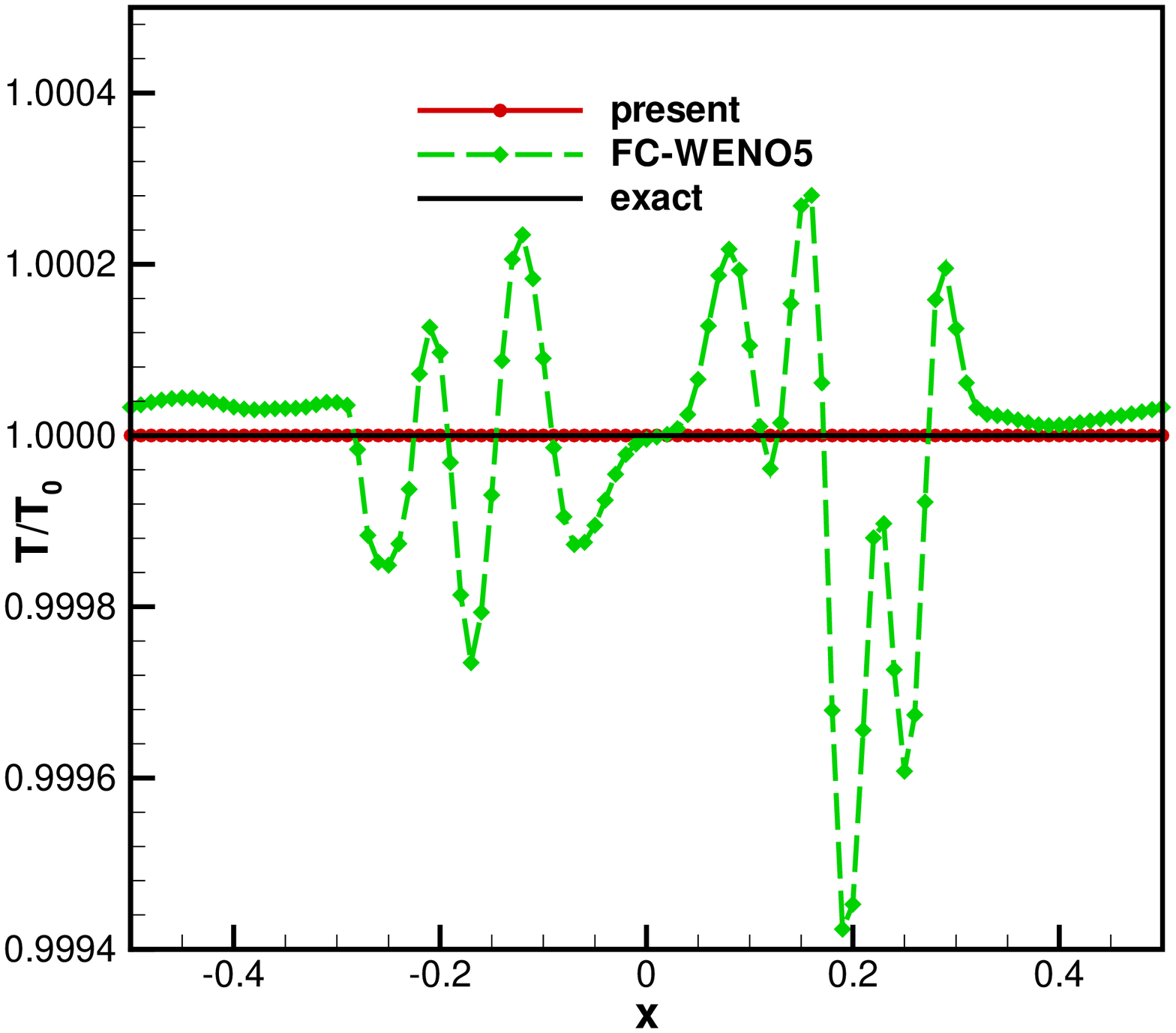}
  \caption{Results of the moving material interface problem ($t=1$).} \label{fig2}
\end{figure}

Fig. \ref{fig2} shows the results. From the figure, we can see that FC-WENO5 generates errors in both the velocity, pressure and temperature profiles. In contrast, a good agreement with the exact solution is obtained when the present algorithm is used to solve the augmented multispecies model. Therefore, we can conclude that the present algorithm is able to maintain the velocity, pressure, and temperature equilibria.

\subsection{Multi-material Sod problem}
Next, we consider the multi-material Sod problem. The initial condition is
\begin{align}
  (\rho,u, p, \gamma, W)=\left\{
                           \begin{array}{ll}
                             (1, 0, 1.0/1.4, 1.4, 28),  & x < 0 \\
                             (0.125, 0, 0.1/1.4, 1.667, 4), & \hbox{otherwise.}
                           \end{array}
                         \right.
\end{align}
The computational domain is $-0.5 \leq x \leq 0.5$. A uniformly distributed grid of 101 points is used, and initial values are fixed on the left and right boundaries of the domain. The final time is $t=0.2$.

\begin{figure}[!ht]
  \centering
  \includegraphics[width=0.48\textwidth]{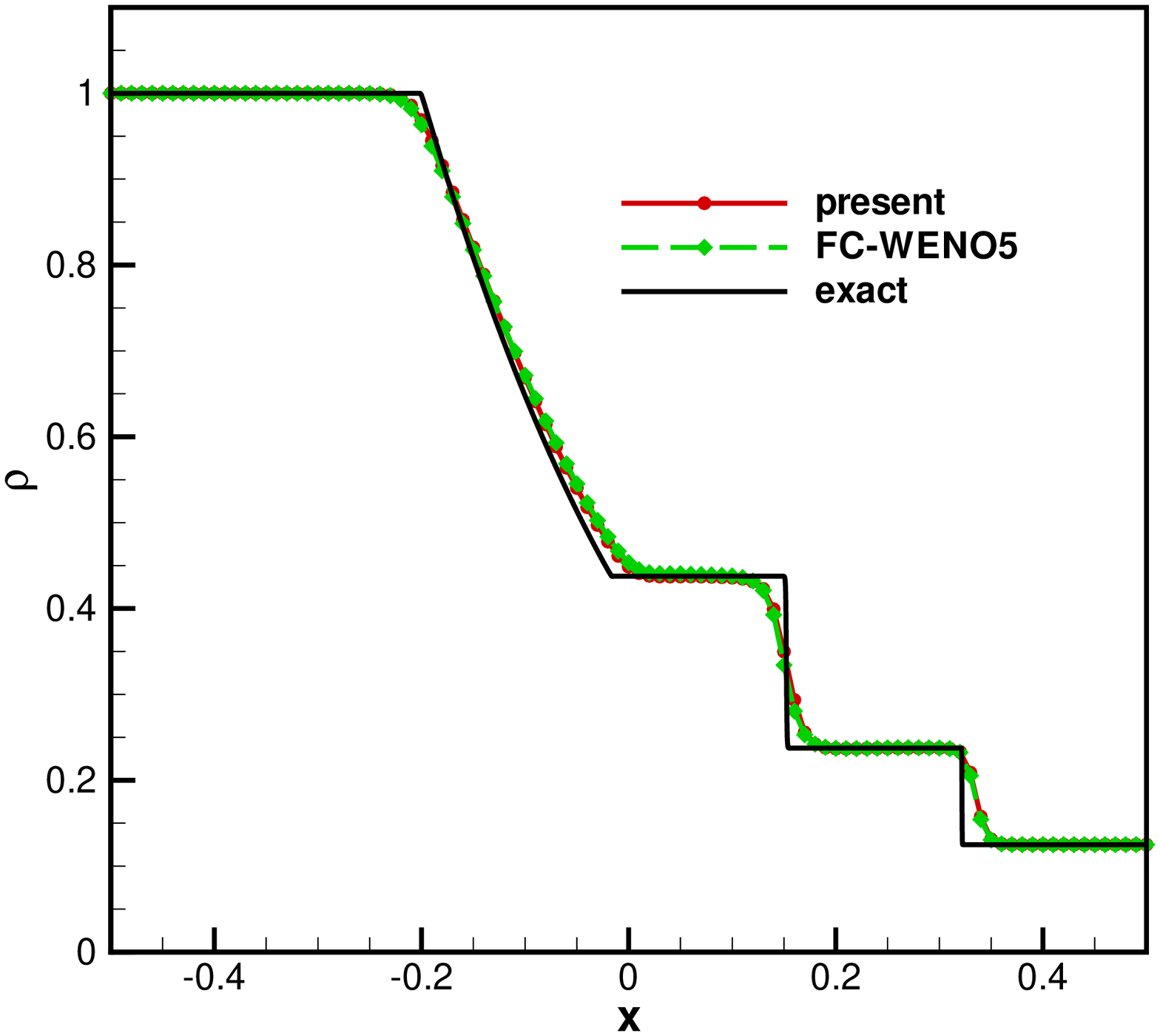}
  \includegraphics[width=0.48\textwidth]{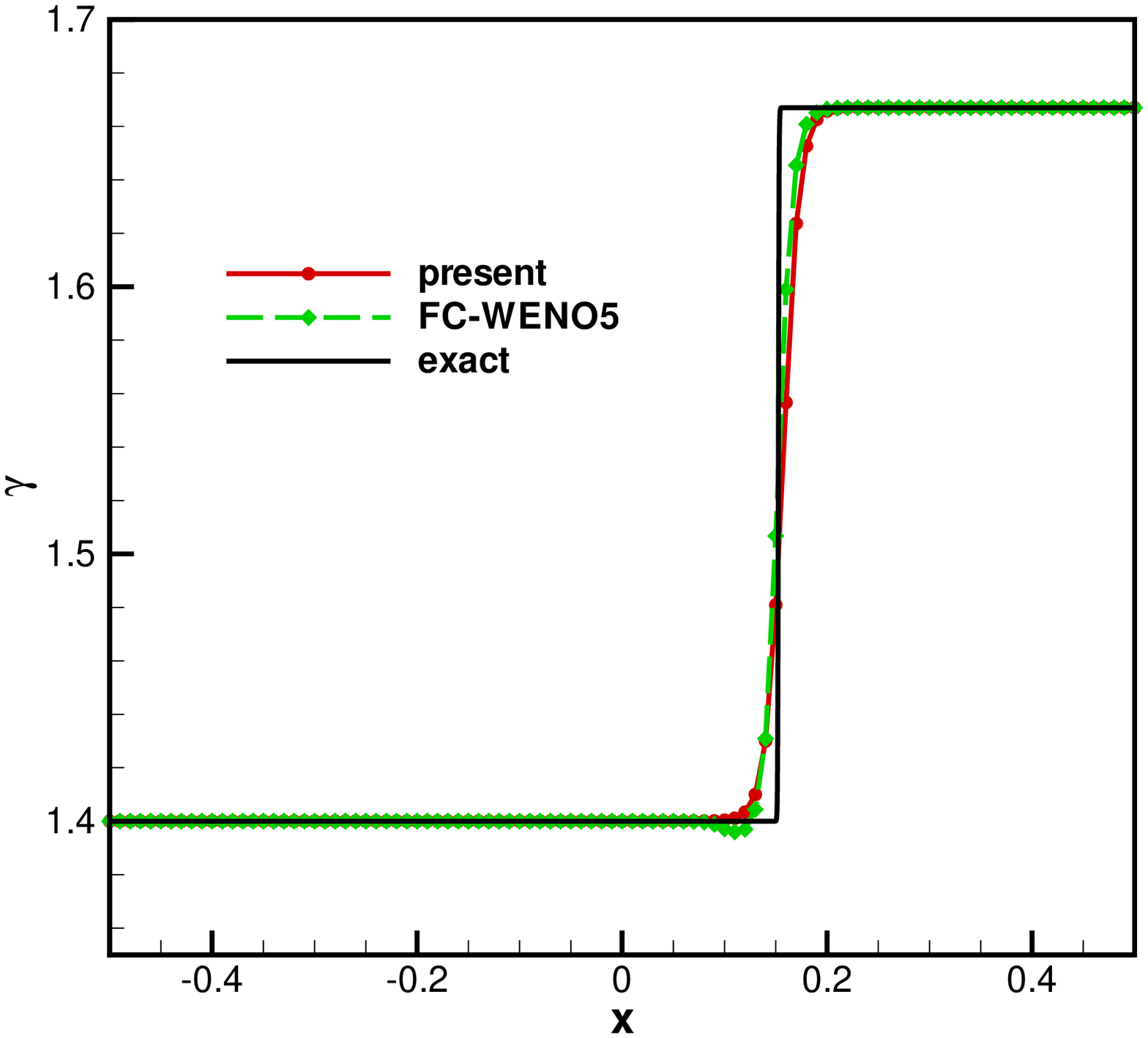}
  \includegraphics[width=0.48\textwidth]{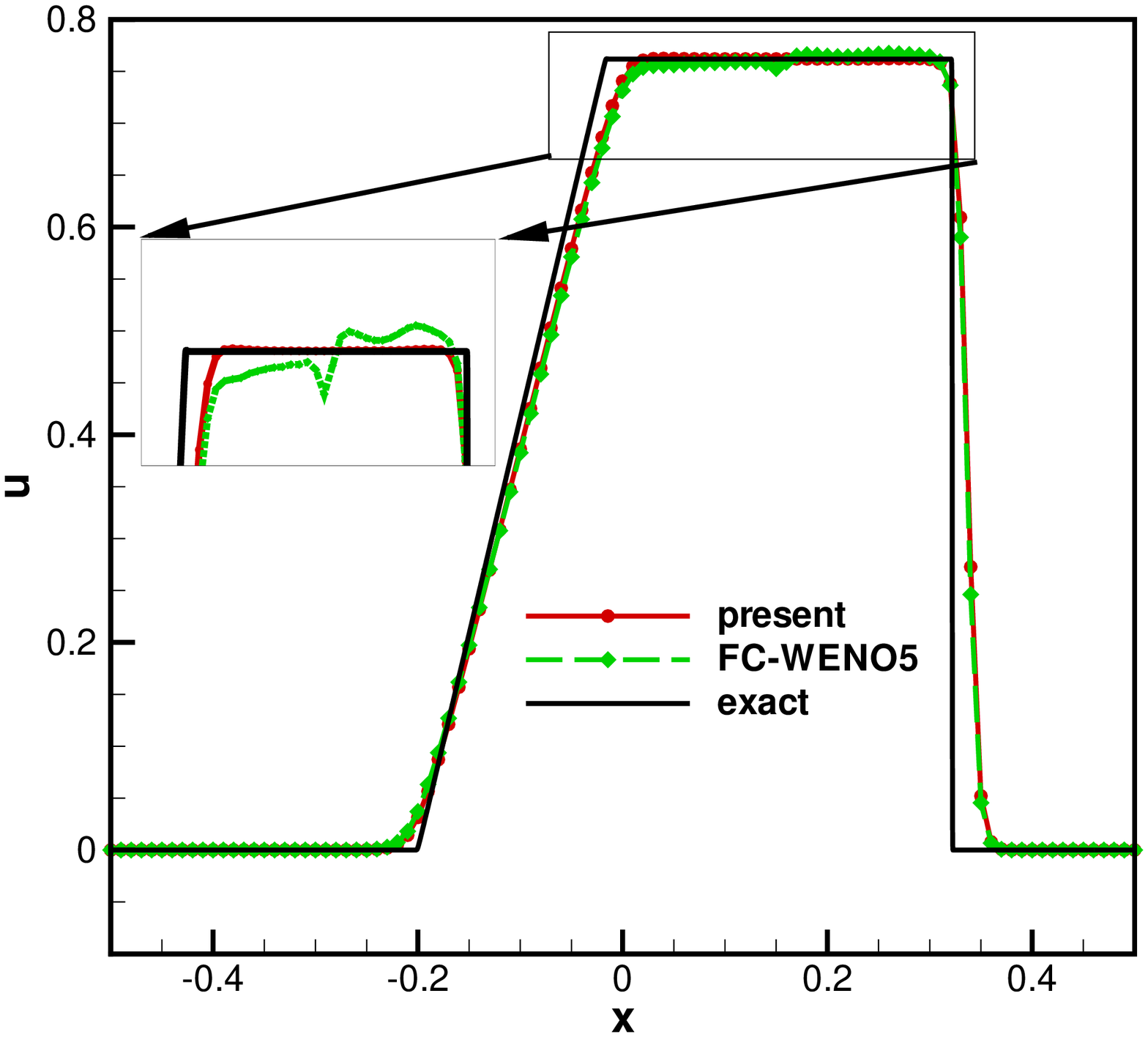}
  \includegraphics[width=0.48\textwidth]{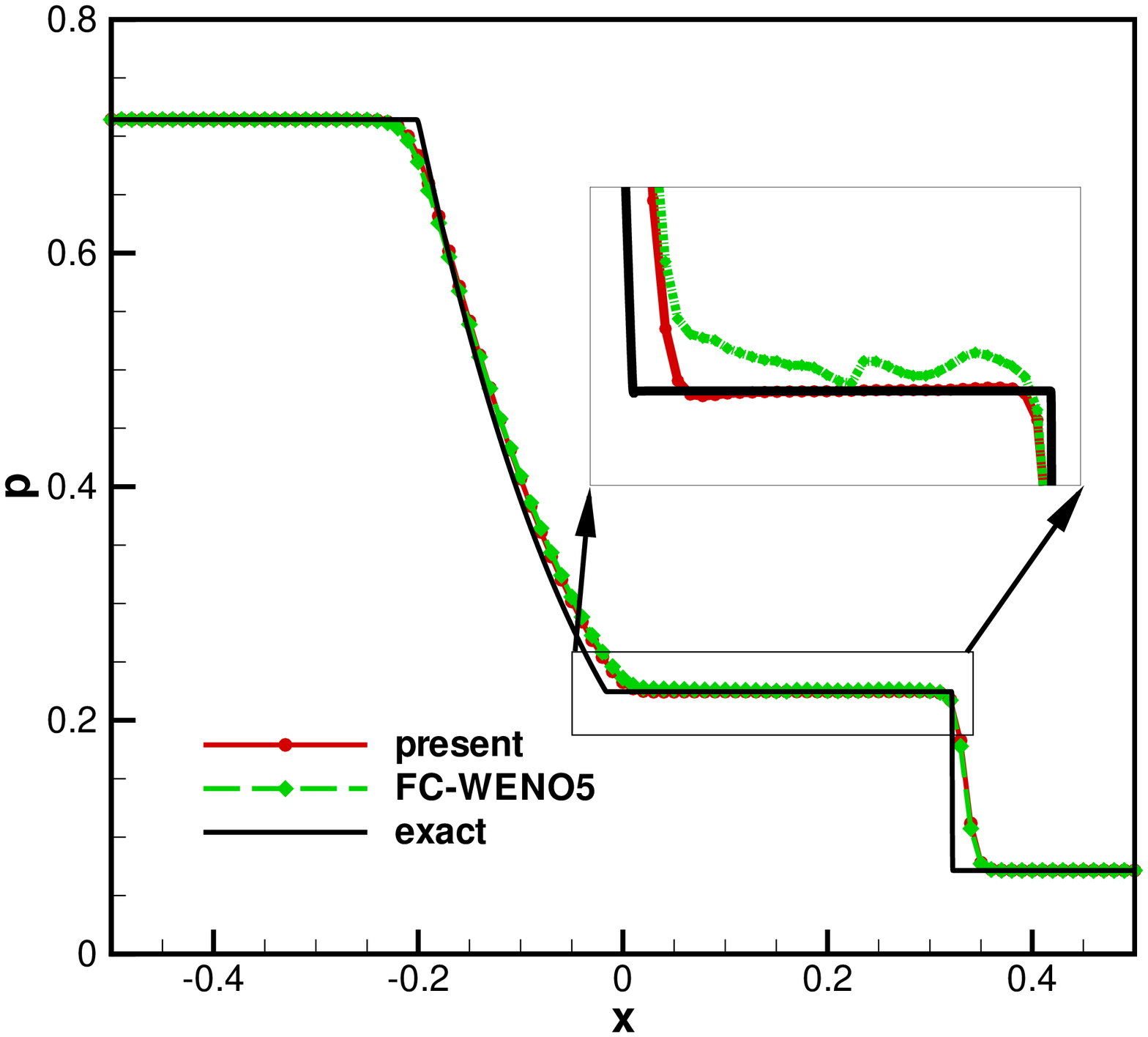}
  \caption{Results of the multi-material Sod problem ($t=0.2$).} \label{fig3}
\end{figure}

Fig. \ref{fig3} shows the final results. Again, FC-WENO5 generates errors in both the velocity and pressure profiles. Furthermore, owing to these errors, other profiles may be affected. For example, there is an overshoot in the back of the jump of $\gamma$ profile. In contrast, a good agreement with the exact solution is noticeable when the present algorithm is used to solve the augmented multispecies model. There are no obvious errors at the interfaces. Moreover, the shock and rarefaction waves are also well captured by the present algorithm. Therefore, the present algorithm can be applied to not only the pure interface problems, but also to problems involving the interaction of interfaces, rarefaction waves, and shock waves.

\subsection{Stiff shock-tube problem}

In this section, the following stiff shock-tube problem \cite{Nonomura2017} is tested:
\begin{align}
  (\rho,u, p, \gamma, W)=\left\{
                           \begin{array}{ll}
                             (1, 0, 100/1.4, 1.4, 28),  & x < 0 \\
                             (1, 0, 0.01/1.4, 1.667, 4), & \hbox{otherwise.}
                           \end{array}
                         \right.
\end{align}
The computational domain is $-0.5 \leq x \leq 0.5$. A uniformly distributed grid of 401 points is used, and the initial values are fixed on the left and right boundaries of the domain. The final time is $t=0.035$.

\begin{figure}[!ht]
  \centering
  \includegraphics[width=0.48\textwidth]{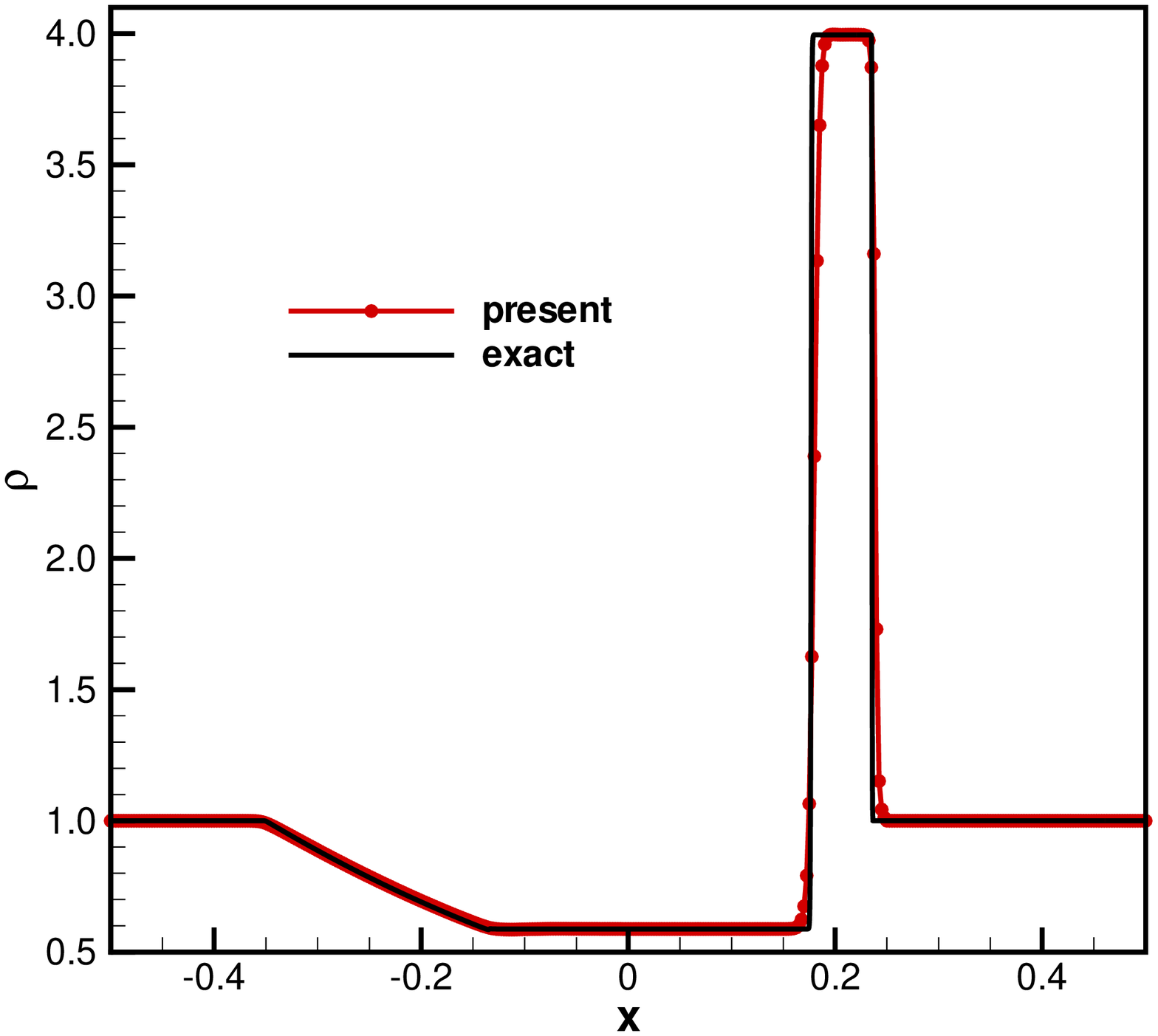}
  \includegraphics[width=0.48\textwidth]{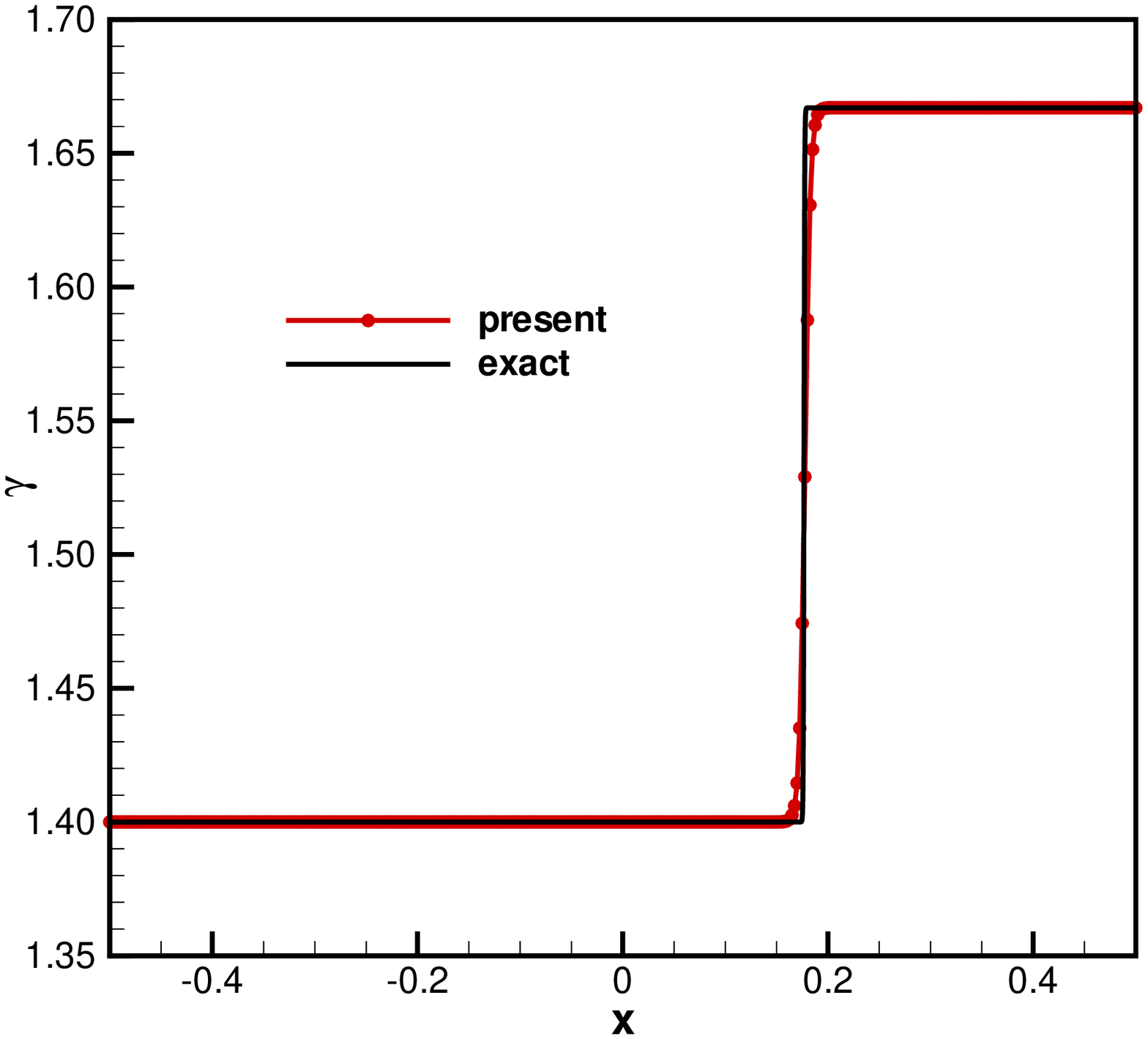}
  \includegraphics[width=0.48\textwidth]{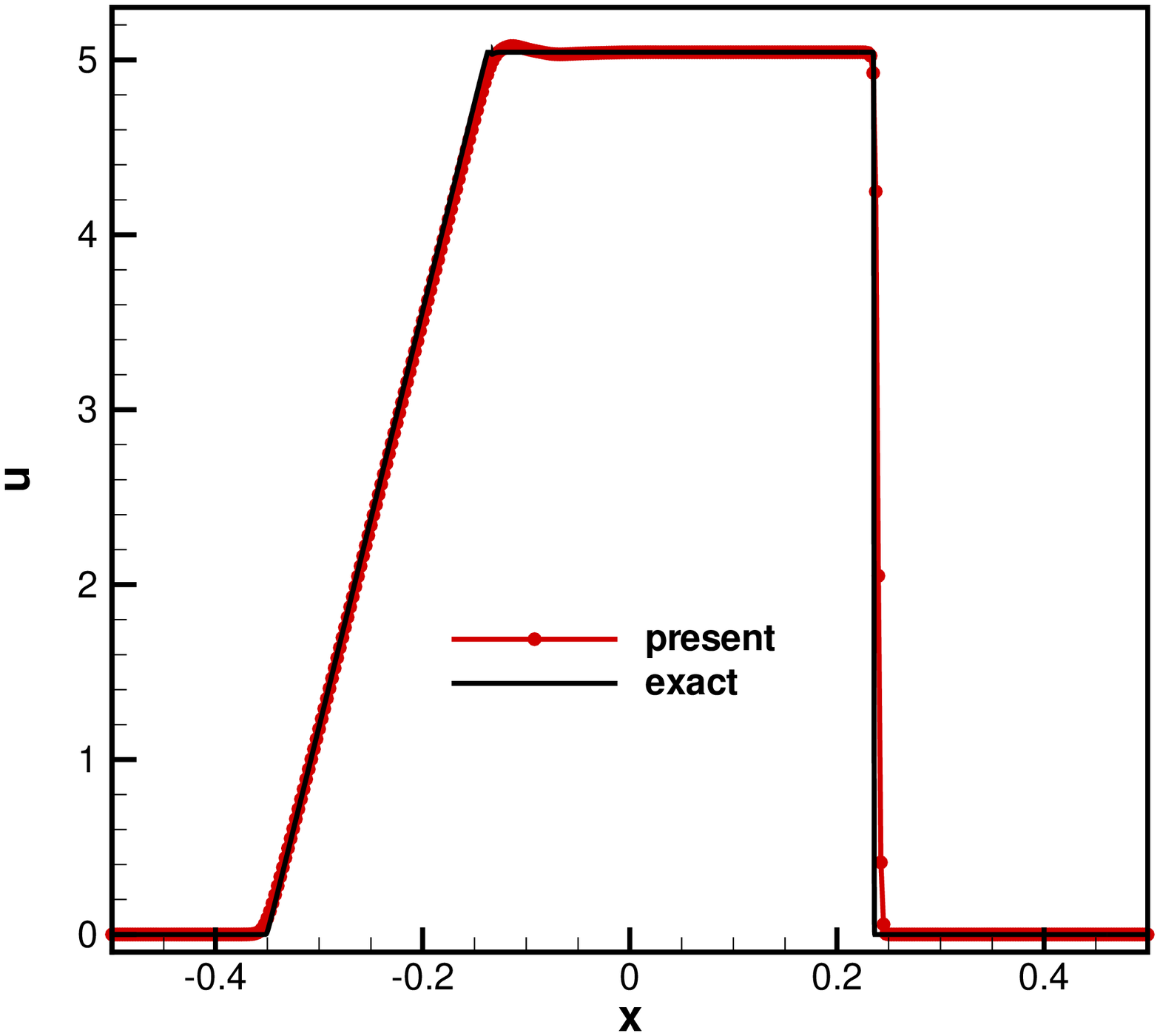}
  \includegraphics[width=0.48\textwidth]{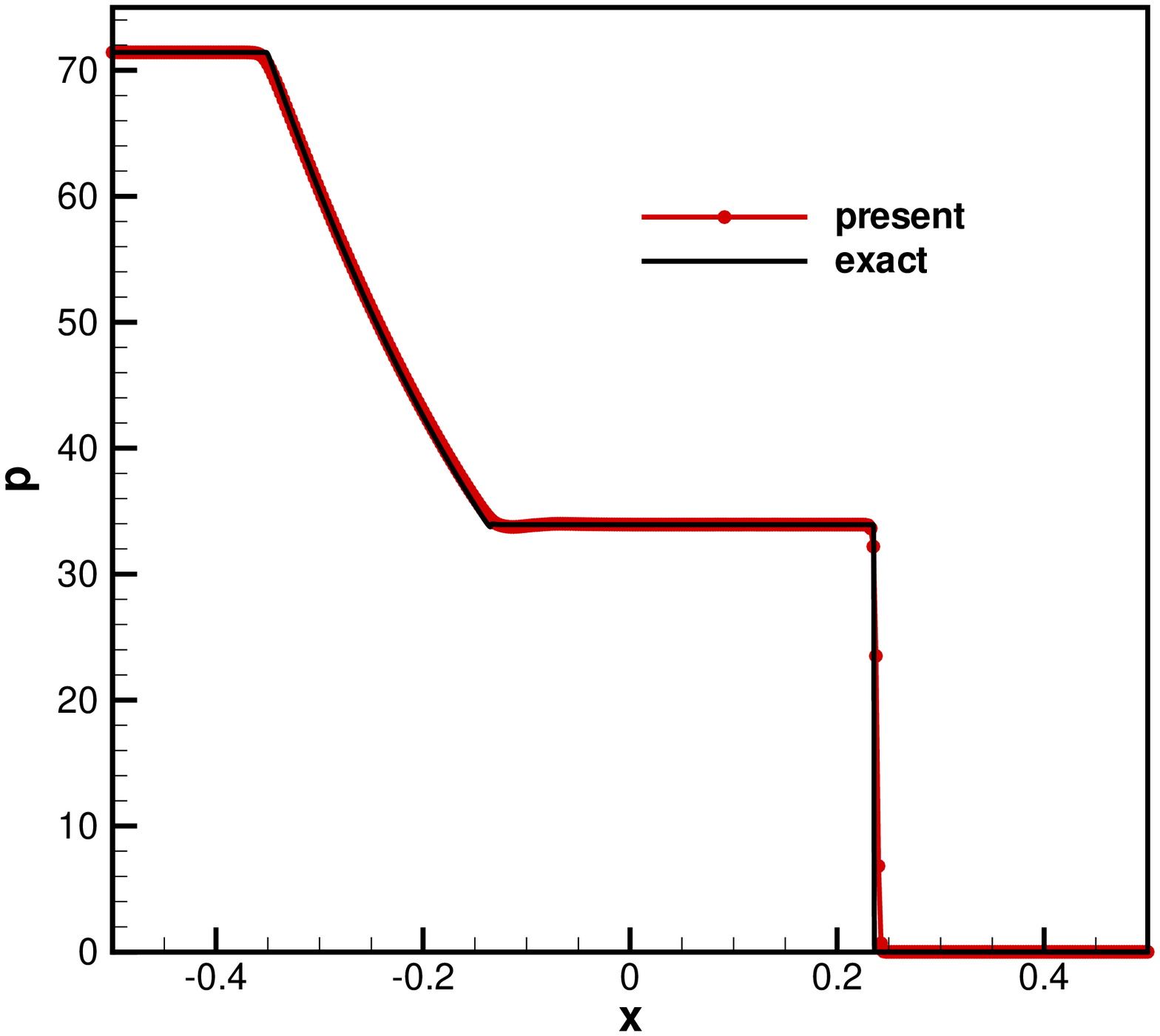}
  \caption{Results of the stiff shock-tube problem ($t=0.035$).} \label{fig4}
\end{figure}

For this problem, FC-WENO5 fails in the early stage of the computation. This phenomenon is the same as that with \cite{Nonomura2017}. The computation breakdown is due to the negative pressure generated by the error for the fully conservative form. In contrast, the present algorithm completes this computation, and Fig. \ref{fig4} shows the final results. From the results, we can observe that a good agreement with the exact solution is obtained. There are no obvious errors at the interfaces, and the strong shock and rarefaction waves are also well captured. These results show that the present algorithm not only has the ability to maintain the equilibria, but also includes sufficient and approximate numerical dissipation for the strong shock and rarefaction waves.

\subsection{Richtmyer--Meshkov instability problem}

Next, the following two-dimensional Richtmyer--Meshkov instability problem is considered:
\begin{align}
  (\rho,u, v, p, \gamma, W)=\left\{
                           \begin{array}{lll}
                             (1.4112, -0.3613, ,0, 1.6272/1.4, 1.4, 28.8),  & x < -0.8 \\
                             (5.04, 0, 0, 1.0/1.4, 1.093, 145.15),  & x < -1.1-0.1 \hbox{cos}(2 \pi y) \\
                             (1, 0, 0, 1.0/1.4, 1.4, 28.8), & \hbox{otherwise.}
                           \end{array}
                         \right.
\end{align}
The computational domain of this problem is $[-8,0] \times [0,1]$, and a uniformly distributed grid of $1025 \times 129$ is used. The outflow boundary conditions are applied at the left and right ends of the domain, and the reflection wall conditions are applied at the bottom and upper ends of the domain.

\begin{figure}[!ht]
  \centering
  \includegraphics[width=0.75\textwidth]{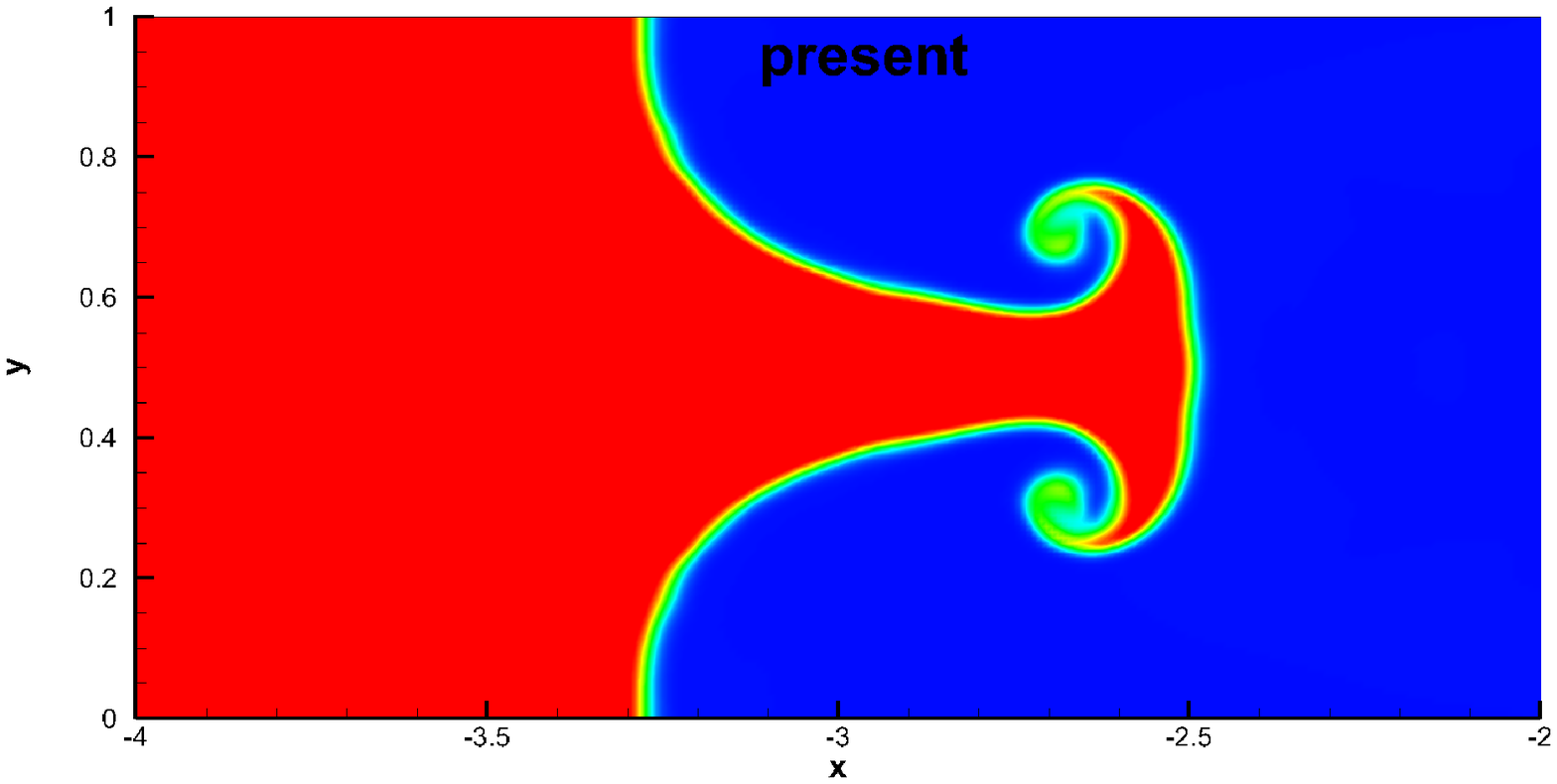}
  \includegraphics[width=0.75\textwidth]{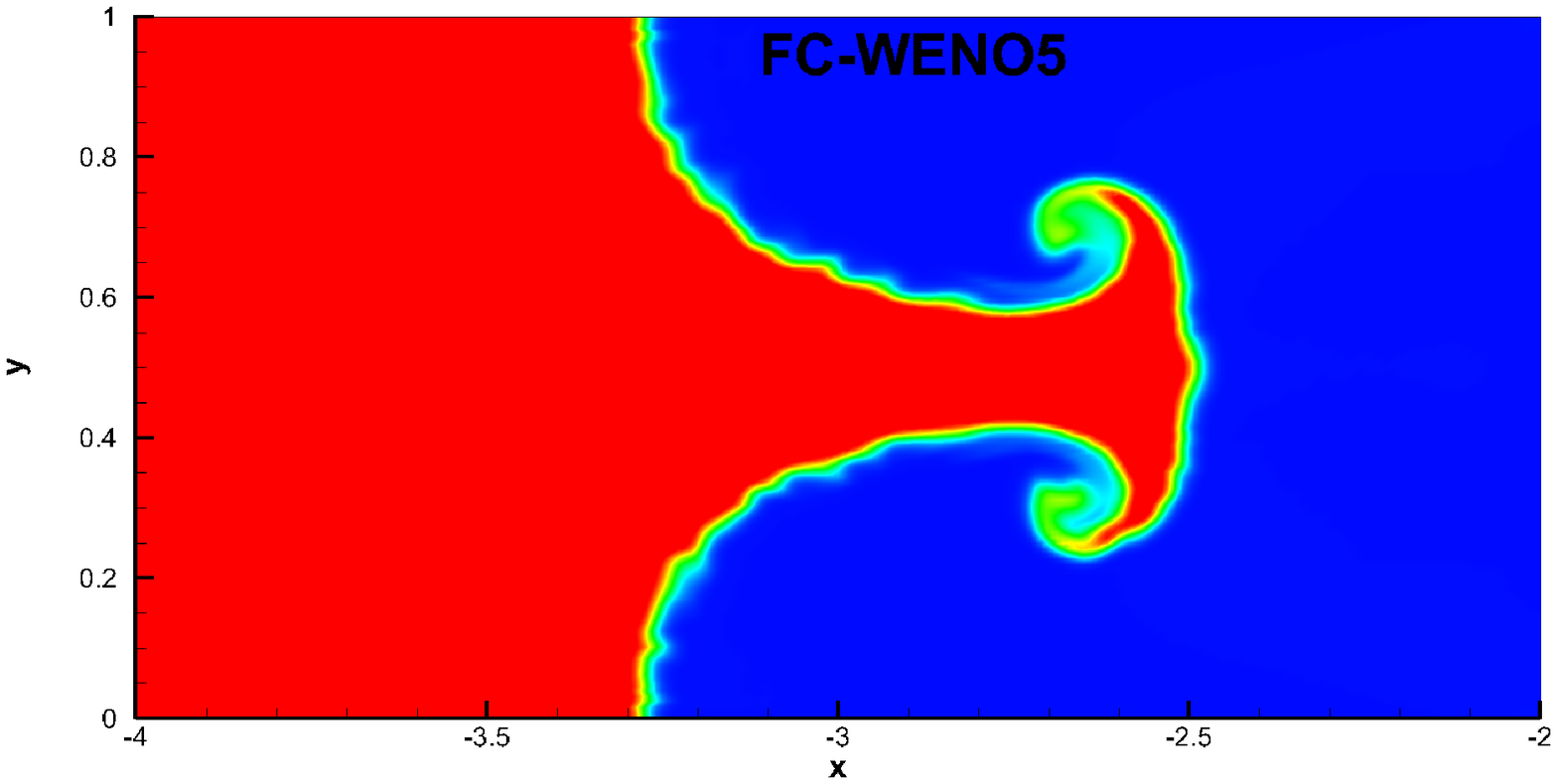}
  \caption{Density fields of the Richtmyer--Meshkov instability problem at $t=8.25$. The contour range is set from 1.5 to 8.5.} \label{fig5}
\end{figure}

Fig. \ref{fig5} shows the interface shape in the late stage at $t=8.25$. From the results, we can see FC-WENO5 cannot maintain a smooth interface shape owing to the generation of the errors in velocity, pressure, and temperature. There are obvious disturbances across the interface, causing it to acquire a sawtooth appearance. In contrast, the present algorithm yields a high-quality result without spurious oscillations. This result confirms that the present algorithm works well for the problems in multi-dimensional systems.

\begin{figure}[!ht]
  \centering
  \includegraphics[width=0.75\textwidth]{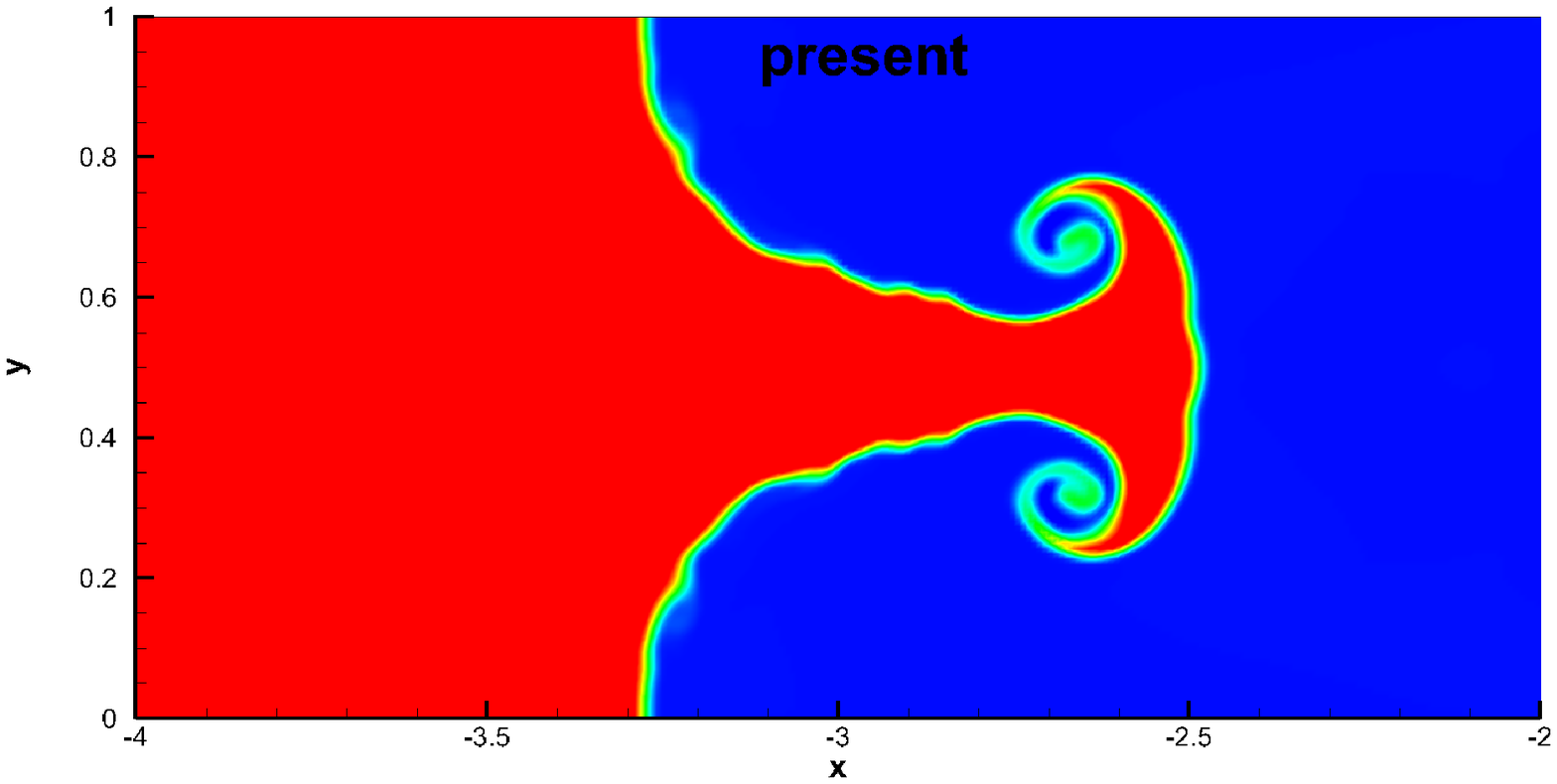}
  \includegraphics[width=0.75\textwidth]{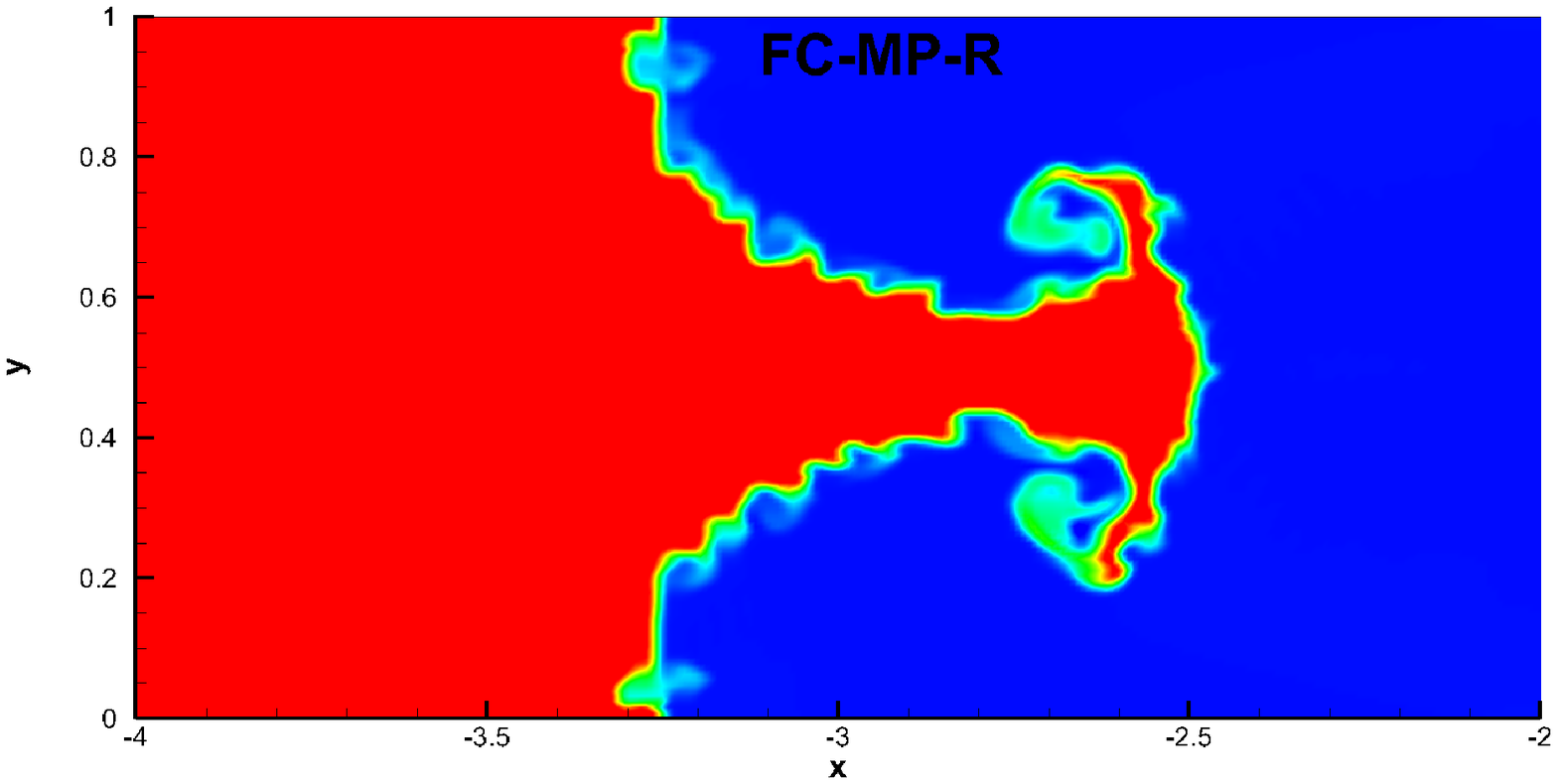}
  \caption{Density fields of the Richtmyer--Meshkov instability problem at $t=8.25$ obtained by the fifth-order MP-R scheme \cite{hezw_mp2}. The contour range is set from 1.5 to 8.5.} \label{fig5_mpr}
\end{figure}

Fig. \ref{fig5_mpr} shows the interface shape at $t=8.25$ obtained by a limiting-type shock-capturing scheme, namely, the improved accuracy MP-R scheme \cite{hezw_mp2} with CFL number 0.4 (a practical upper bound).

From the results, we can observed that the fully conservative formulation solved by the fifth-order MP-R scheme (FC-MP-R) produces severe errors for the velocity, pressure, and temperature that drastically affect the creditability of the result. However, these errors can be remarkably reduced by the present algorithm. This result indicates that the present algorithm can be directly applied to any conservative shock-capturing scheme.

\subsection{Shock-bubble interaction problem}

Finally, the following two-dimensional shock-R22-bubble problem \cite{Nonomura2017} is considered:
\begin{align}
  (\rho,u, v, p, \gamma, W)=\left\{
                           \begin{array}{lll}
                             (1.3764, -0.3336, ,0, 1.5698/1.4, 1.4, 28.8),  & x \geq 1 \\
                             (3.153, 0, 0, 1.0/1.4, 1.249, 90.82),  & \sqrt{x^2+y^2} < 0.5 \\
                             (1, 0, 0, 1.0/1.4, 1.4, 28.8), & \hbox{otherwise.}
                           \end{array}
                         \right.
\end{align}
The computational domain of this problem is $[-3.5,3] \times [-0.89,0.89]$, and a uniformly distributed $651 \times 179$ grid is used. The boundary conditions of this problem are the same as the previous case.

\begin{figure}[!ht]
  \centering
  \includegraphics[width=0.49\textwidth]{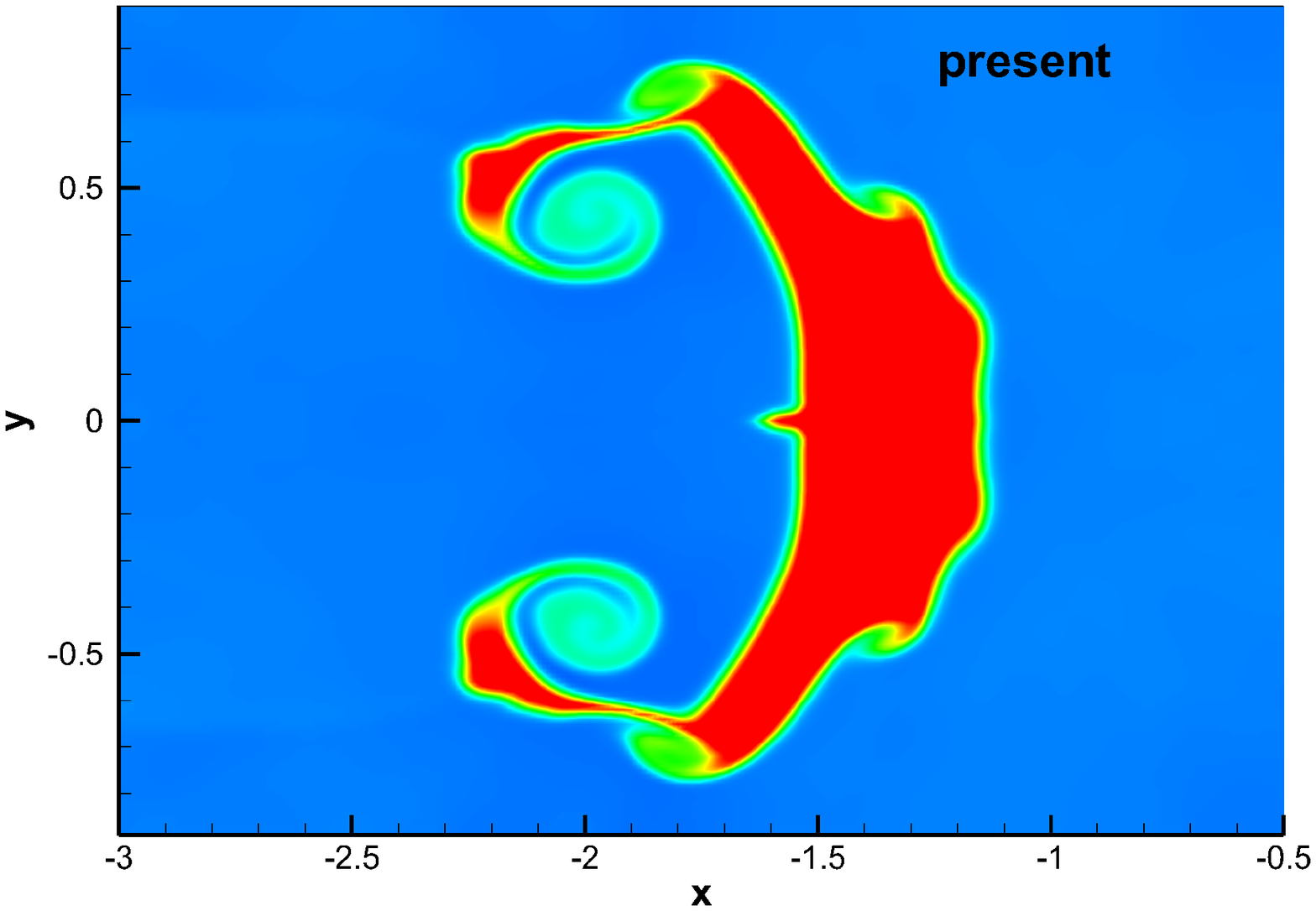}
  \includegraphics[width=0.49\textwidth]{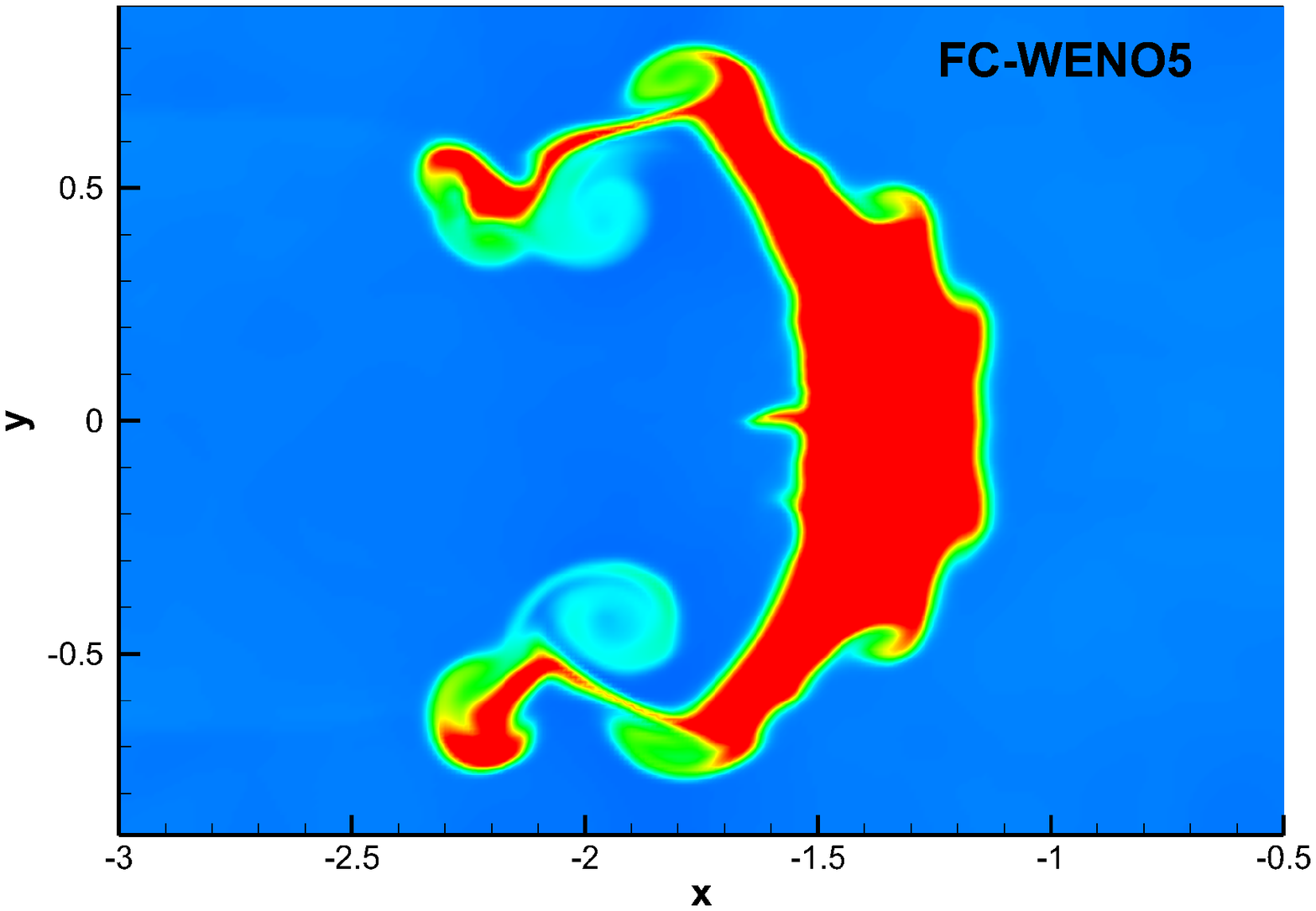}
  \caption{Density fields of the shock-bubble interaction problem at $t=7.337$. The contour range is set from 1 to 4.} \label{fig6}
\end{figure}

\begin{figure}[!ht]
  \centering
  \includegraphics[width=0.49\textwidth]{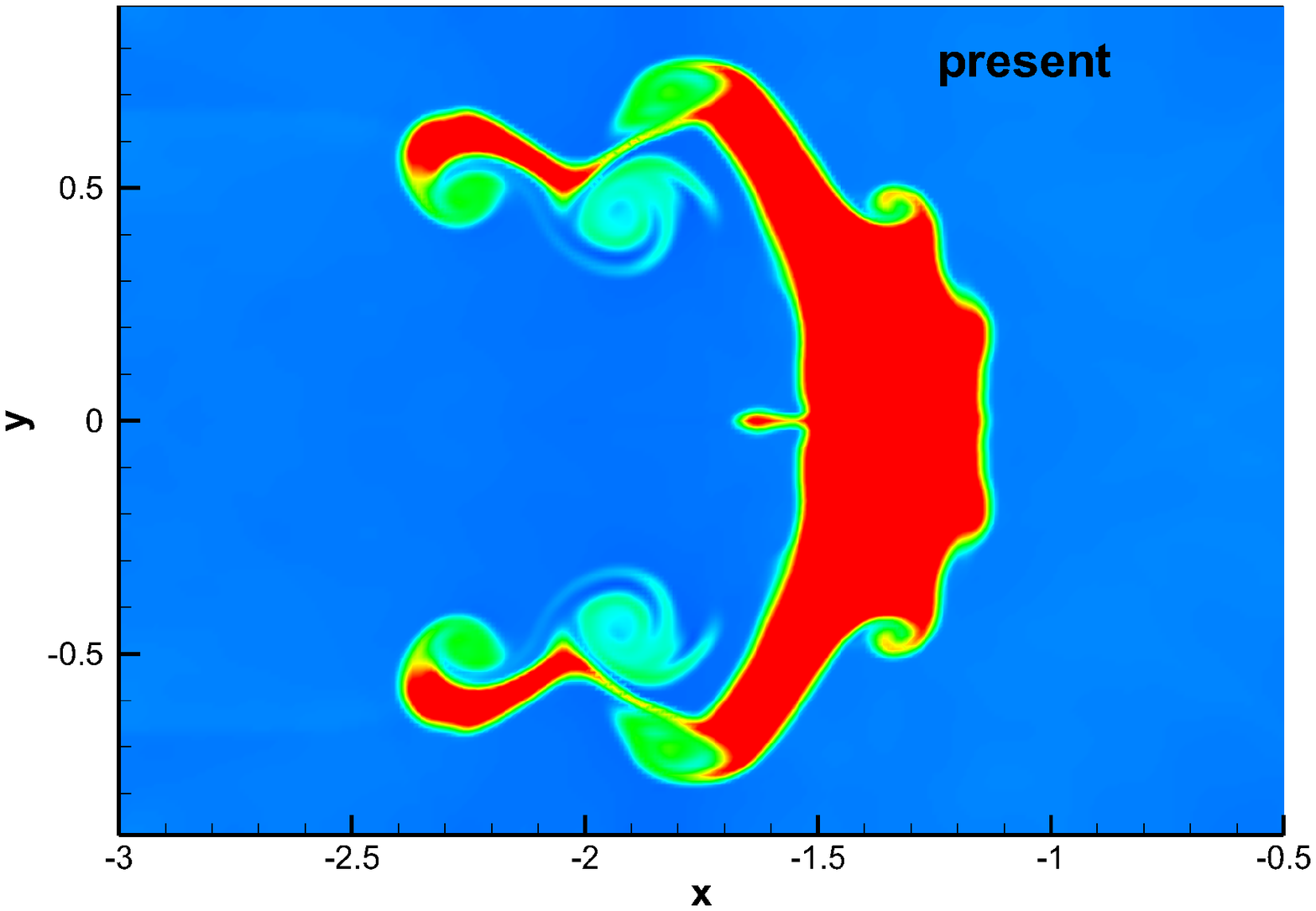}
  \includegraphics[width=0.49\textwidth]{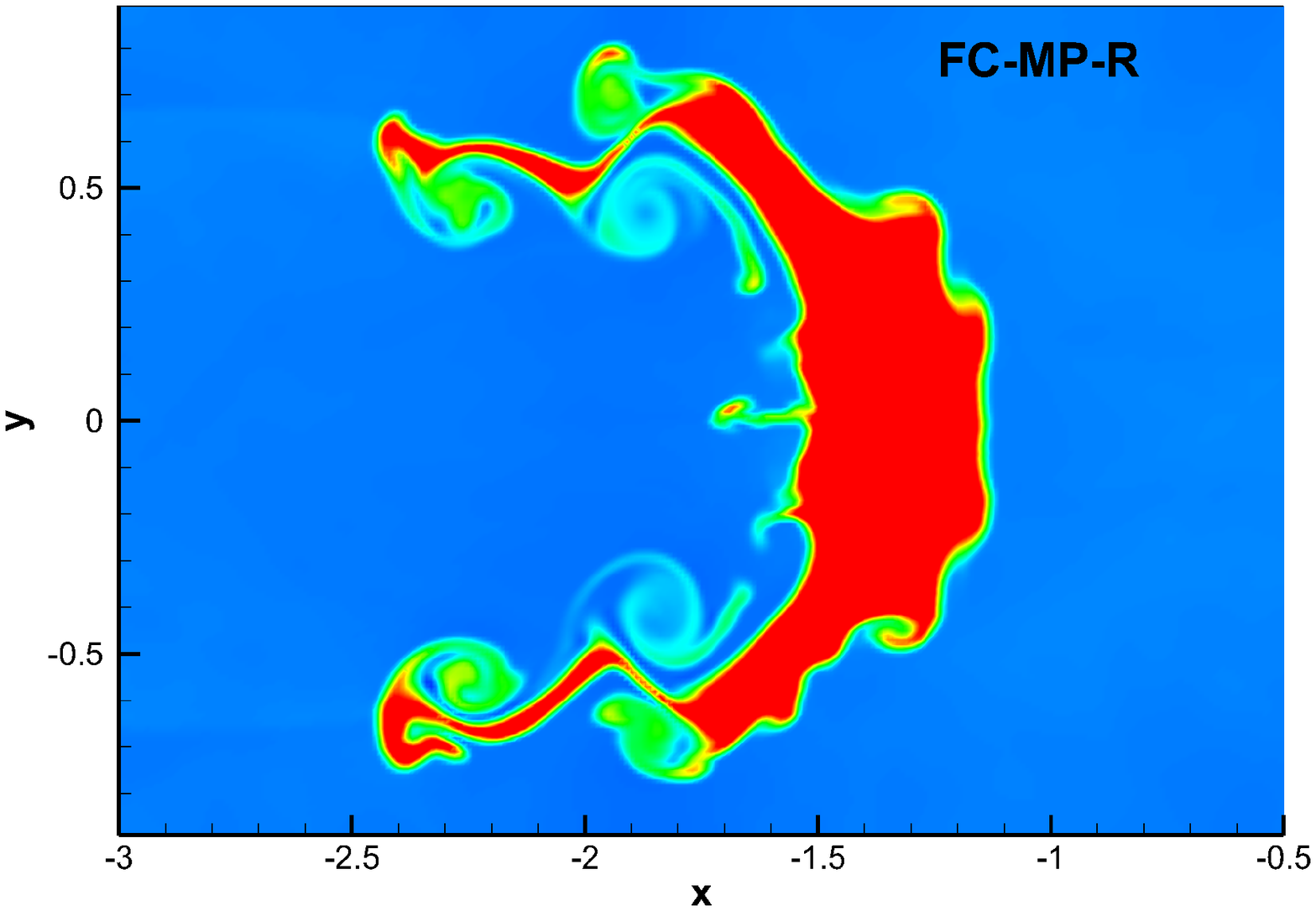}
  \caption{Density fields of the shock-bubble interaction problem at $t=7.337$ obtained by the fifth-order MP-R scheme. The contour range is set from 1 to 4.} \label{fig6_mpr}
\end{figure}

Fig. \ref{fig6} shows the interface shape at $t=7.337$, whereas Fig. \ref{fig6_mpr} displays the same obtained by the fifth-order MP-R scheme. As before, the fully conservative formulation cannot maintain a smooth interface shape. Furthermore, the symmetry of this result is slightly broken owing to these errors in the velocity, pressure, and temperature. However, the present algorithm can still yield a high-quality result without spurious oscillations. These results again validate the conclusions stated above.

\section{Conclusions}
The present work mainly focuses on designing a general consistent finite difference algorithm for non-conservative numerical models of multi-material gas flows (when the specific heats ratio is variable). First, we provide a detailed review of the previous methods for the consistent implementation of a high-order characteristic flux-split-based FDM for such flows, and reveal why special treatments (e.g., common discretization of the genuinely nonlinear fields \cite{hezw2016} and use of the split form of a nonlinear WENO scheme \cite{Nonomura2017}) are required in these methods. Based on this analysis, we rewrite the non-conservative term as a conservative term with a source term containing the velocity divergence, and propose a general framework that is found to have the ability of maintaining the velocity, pressure, and temperature equilibria. Furthermore, the consistent discretization form of the velocity divergence in the source term is also determined by imposing a new criterion wherein a multi-fluid algorithm should have the ability of maintaining a pure single-fluid. Based on these works, a new general algorithm that does not include any special treatment is proposed, and is validated by several one- and two- dimensional numerical tests.

This new consistent algorithm, for the first time, provides a general high-order FDM to discretize a source term containing the velocity divergence, and does not require any special treatment. A five-equation model has the same type of source term containing the velocity divergence \cite{Kapila2001, Murrone2005, Abgrall2006fiveeqns}; therefore, it is expected that this new algorithm can be applied to this model. This work is ongoing, and will be reported in the coming future.

\section*{Acknowledgement}
This work was supported by NSFC under Grant Nos. 201702002.

\section*{References}

\bibliography{mybibfile,mybibfile2}

\end{document}